\documentclass[a4paper,aps,prd,preprint,tightenline]{revtex4-1}

\usepackage{color}
\usepackage{amsmath}
\usepackage{accents}
\usepackage{amssymb}
\usepackage{graphicx}
\usepackage{marginnote}

\usepackage[colorlinks]{hyperref} 
\hypersetup{
	citecolor=blue,
	linkcolor=blue,
	urlcolor=blue
}

\newcommand{\teal}[1]{\textcolor{teal}{#1}}
\newcommand{\defineText}[2]{\newcommand{#1}[0]{\text{#2}}}

\newlength{\dhatheight}
\newcommand{\doublehat}[1]{%
    \settoheight{\dhatheight}{\ensuremath{\hat{#1}}}%
    \addtolength{\dhatheight}{-0.35ex}%
    \hat{\vphantom{\rule{1pt}{\dhatheight}}%
    \smash{\hat{#1}}}}
\defineText{\MV}{MV}
\defineText{\GV}{GV}
\defineText{\MeV}{MeV}
\defineText{\GeV}{GeV}
\defineText{\TeV}{TeV}
\defineText{\cm}{cm}
\defineText{\km}{km}
\defineText{\kpc}{kpc}
\defineText{\sr}{sr}
\defineText{\LIS}{LIS}
\defineText{\TOA}{TOA}

\usepackage{xcolor}
\usepackage{tgtermes} 

\makeatother

\usepackage{babel}

\usepackage{comment}

\renewcommand{\figurename}{FIG.}

\newcommand{\fig}[1]{Fig.~\ref{#1}}
\renewcommand{\tablename}{TAB.} 

\newcommand{\tab}[1]{Tab.~\ref{#1}}
\newcommand{\beq}[1]{\begin{equation}\label{#1}}
	\newcommand{\eeq}{\end{equation}}
\newcommand{\bea}[1]{\begin{eqnarray}\label{#1}}
	\newcommand{\eea}{\end{eqnarray}}
	
\usepackage{tcolorbox}
\newcounter{reviewer}
\newcounter{point}[reviewer]
\renewcommand{\thepoint}{\thereviewer.\arabic{point}}

\begin{document}
\title{Constraints on evaporating primordial black holes from \\ the AMS-02
positron  data}
\author{Jia-Zhi Huang}
\affiliation{Institute of Theoretical Physics, 
	Chinese Academy of Sciences, Beijing 100190, China. \\
	University of Chinese Academy of Sciences, Beijing, 100190, China.}
\author{Yu-Feng Zhou}
\affiliation{Institute of Theoretical Physics, 
	Chinese Academy of Sciences, Beijing 100190, China. \\
	University of Chinese Academy of Sciences, Beijing, 100190, China.}
\affiliation{School of Fundamental Physics and Mathematical Sciences, 
	Hangzhou Institute for Advanced Study, UCAS, Hangzhou 310024, China. \\
	International Centre for Theoretical Physics Asia-Pacific, Beijing/Hangzhou,
China.}
\date{\today}
\begin{abstract} 
	Cosmic-ray (CR) positrons are relatively rare due to their secondary origin and thus sensitive to exotic contributions. Primordial black holes (PBHs) with masses above $\sim 5\times10^{14}\,\mathrm{g}$ can be stable sources of CR positrons due to Hawking radiation. The energies of the evaporated positrons can increase significantly through scattering with the Galactic random magnetic fields during the propagation in the Galaxy, which is a generic feature in  diffusive re-acceleration CR propagation models. We show that in well-constrained diffusive re-acceleration models, a significant portion of CR positron flux can enter the energy region of $\mathcal{O}(\text{GeV})$, and can be constrained by the current AMS-02 data. As an example, we show that in the
	{\tt Galprop+Helmod} model for CR propagation in the Galaxy and heliosphere, an upper limit of $f_{\text{PBH}}\lesssim2.15\times 10^{-4}$ at PBH mass $2\times 10^{16}$~g can be obtained, which improve the previous constraints from the Voyager CR all-electron data by around an order of magnitude.
\end{abstract}
\maketitle
\section{Introduction} \label{sec:intro} 
Astrophysical and cosmological observations suggest that the dominant component of the matter in the present Universe is in the form of non-luminous dark matter (DM). In many particle physics models, DM consists of  a new type of elementary particle beyond the Standard Model (SM) of particle physics, which may participate in non-gravitational interactions with baryonic matter, 
	such as weakly interacting massive particles (WIMPs), 
	sterile neutrinos 
	and QCD axions (for recent reviews of these particle candidates, see e.g.~\cite{Bauer:2017qwy,Abazajian:2012ys,Marsh:2015xka}). 
Despite great experimental efforts in recent decades, 
so far there is no confirmed signals of particle DM from direct, 
indirect and collider DM search experiments.

Primordial black hole (PBH) is an alternative DM candidate which does not
require any new physics beyond the SM~\cite{Hawking:1971ei,Carr:1974nx}. 
PBHs are believed to have formed after the inflation~\cite{Tashiro:2008sf,Germani:2018jgr,Kannike:2017bxn,Carr:2017jsz,Carr:2017edp},
and subsequently evolved through accretion, mergers, and Hawking radiation. 
Depending on the time of formation, the masses of PBHs can vary in a
large range. In general, the initial mass $M_{\text{PBH}}$ of a PBH should be
close to the Hubble horizon mass at the production time $t$,
$M_{\text{PBH}}\sim c^{3}\,t/G\simeq10^{15}(t/10^{-23}\,\text{s})\,\text{g}$,
where $c$ is the speed of light and $G$ is the Newton constant. For two
typical formation time $t$ of the Planck time $t\sim10^{-43}\,\mathrm{s}$ and
the time just before the big-bang nucleosynthesis (BBN) $t\sim1\,\mathrm{s}$, the initial PBH
mass are respectively around $10^{-5}\,\mathrm{g}$ and $10^{5}\,M_{\odot}$, where $M_{\odot}$ is the mass of the Sun. 

PBH can comprise all or a fraction of the DM. The energy fraction of PBHs
relative to that of whole DM is defined as $f_{\text{PBH}}\equiv\Omega_{\text{PBH}}/\Omega_{\text{DM}}$,
where $\Omega_{\text{DM}}$ and $\Omega_{\text{PBH}}$ are the energy
density parameters of DM and PBHs relative to the critical density of the present
Universe, respectively. 
For heavy PBHs with $M_{\text{PBH}}\gg10^{17}\text{g}$,
the value of $f_{\text{PBH}}$ can be constrained by the gravitational
effects from PBHs such as microlensing~\cite{MACHO:2000qbb,Wyrzykowski:2010mh,Wyrzykowski:2011tr,Macho:2000nvd,Griest:2013aaa,CalchiNovati:2013jpj,Griest:2013esa},
dynamical constraint from globular clusters, galaxy disruption and other observables~\cite{Koushiappas:2017chw,Monroy-Rodriguez:2014ula,Carr:2019bel,MUSE:2020qbo}
(for recent reviews, see e.g.~\cite{Carr:2020gox,Carr:2020xqk}). 

Light PBHs are expected to emit SM particles through Hawking radiation~\cite{Hawking:1974rv}.
PBHs in the mass range from $10^{13}-10^{17}\,\text{g}$ are expected
to emit SM particles with typical energies from a few GeV down to
a few hundreds of keV. PBHs lose their masses through Hawking radiation at a rate $\text{d}M_\text{PBH}/\text{dt} \propto M_\text{PBH}^{-2}$~\cite{Baldes:2020nuv,Cheek:2021odj}, which suggests that heavier PBHs evaporate slower.
%
%
It has been shown that PBHs with masses $M_{\text{PBH}}\gtrsim 5\times10^{14}\text{ g}$
have lifetimes larger than the age of the Universe, and can be considered as stable sources of photons and cosmic-ray (CR) particles in the Galaxy~\cite{MacGibbon:1991tj}.
This type of evaporating PBHs can be searched by current space-borne experiments.
For PBHs in this mass region, the value of $f_{\mathrm{PBH}}$ can
be constrained by the data of extragalactic and 
galactic diffuse $\gamma$-rays~\cite{Carr:2009jm,Carr:2016hva,Arbey:2019vqx,Laha:2020ivk}, 
CMB~\cite{Auffinger:2022khh},
neutrinos~\cite{Dasgupta:2019cae,Wang:2020uvi,Bernal:2022swt,Liu:2023cqs},
CR electrons~\cite{Boudaud:2018hqb},
511 keV gamma-ray lines~\cite{Dasgupta:2019cae,Laha:2019ssq,Keith:2021guq,luque2024refininggalacticprimordialblack},
21-cm radio signals~\cite{Mittal:2021egv} 
and CR antiprotons~\cite{Maki:1995pa,Barrau:2001ev}, etc.

The measured CR fluxes can also place importance constraints on $f_\text{PBH}$. 
Recently, low-energy CR all-electron ($e^{+}+e^{-}$) flux data from Voyager-1
have been used to set constraints on $f_{\text{PBH}}$~\cite{Boudaud:2018hqb}.
The obtained limits turned out to be competitive with that derived from
extragalactic $\gamma$-rays. As Voyager-1 is now outside the heliopause, the
electron flux measured by Voyager-1 can be considered as the true local
interstellar (LIS) flux, and the derived constraints are expected to be
relatively robust against the influence of the solar activity, the so called solar modulation effect.
Note, however, that although the low-energy electron data from Voyager-1 can be considered as the true LIS flux, the theoretical prediction for CR flux from PBH evaporation involves a  number of parameters for the CR propagation within the Galaxy, such as the diffusion coefficient, re-acceleration coefficient and convection velocity. These parameters are determined  through fitting to the CR data (e.g the Boron to Carbon flux ratio, B/C) measured at the top of the atmosphere (TOA) deep inside the heliosphere, which are strongly affected by the solar activities.  Thus, the constraints on the PBH fraction from the Voyager data
are \textit{inevitably} affected by the solar modulation effect.
Actually, the LIS fluxes measured by Voyager-1, 2 are more useful in improving the  modeling and calibrating of the solar modulation effect itself. In order to derive robust constraints on  exotic contributions, it is necessary to consider both the  LIS and TOA CR flux data simultaneously and consistently calculate the CR propagation in the Galaxy and heliosphere, as the parameters of the two processes are strongly correlated.
%

In this work we explore the possibility of using CR positron flux to constrain the abundance of PBHs. CR positrons are  believed to be of secondary origin and relatively rare. Thus they are sensitive to  exotic contributions. The low-energy CR positron flux can be used to set constraints on exotic contributions as they are roughly consistent with the expected backgrounds (see, e.g.~\cite{Jin:2017iwg}).
Although the initial energy of the CR positrons from the PBHs with mass above $5\times10^{14}~\text{g}$ should be well below $\mathcal{O}(10)$~MeV, the energies of final positrons can increase significantly through scattering with the Galactic random magnetic fields during the propagation process~\cite{Berezinsky:1990qxi}, which is a typical feature in many diffusive re-acceleration CR propagation models~\cite{Trotta:2010mx,Jin:2014ica,Yuan:2018lmc,Boschini:2017fxq,Boschini:2019gow,Boschini:2020jty,DeLaTorreLuque:2021yfq,Luque:2021nxb}.
%
It has been shown that CR propagation plays an important role in constraining exotic contributions from CR data (see e.g.~\cite{DelaTorreLuque:2023huu,DelaTorreLuque:2023nhh,DelaTorreLuque:2023olp,DelaTorreLuque:2023cef}).
In this work, we show that in some well-constrained diffusive re-acceleration models, a significant portion of CR positrons can enter the GeV region, and can be constrained by the current AMS-02 experiment which can measure the CR positron flux with kinetic energy down to 0.6~GeV.
We use the state-of-the-art models for CR propagation in the Galaxy 
(using the  numerical code ${\tt{Galprop}}$
\cite{Strong:1998pw,Moskalenko:2001ya,Strong:2001fu,Moskalenko:2002yx,Ptuskin:2005ax}) and 
in the heliosphere (using  the numerical solutions of the Parker equation 
based on the ${\tt {Helmod}}$ code
\cite{Bobik:2011ig,Bobik:2016,Boschini:2017gic,Boschini:2019ubh,Boschini:2022jwz}
and the analytical force-field approximation).
In calculating the CR fluxes from PBH evaporation, we use the ${\tt{BlackHawk}}$~\cite{Auffinger:2022sqj} code,  which includes both the primary and secondary particle components.
The results show that  for typical diffusive re-acceleration models the AMS-02 positron data can provide very stringent limits, which can be  stronger than the previous constraints derived from the Voyager all-electron flux by around an order of magnitude. 

The remaining part of this paper is organized as follows. 
In section~\ref{sec:phb}, we give a brief overview on the positron energy spectrum from PBH evaporation. In section~\ref{sec:propagation}, we discuss the CR propagation in the Galaxy and the models for solar modulation. The constraints on PBH abundances from the AMS-02 positron data are discussed in section~\ref{sec:constraints}
for the cases without and with background included. The results of this work is summarized in section~\ref{sec:conclusion}.

\section{Evaporation of primordial black holes}\label{sec:phb}
 In this work, unless otherwise stated, we adopt the natural
system of units with $\hbar=k_{\text{B}}=c=1$, where $\hbar$ is
the reduced Plank constant, $k_{\text{B}}$ is the Boltzmann constant,
and $c$ is the speed of light. We consider a simple scenario where
the spin of PBHs can be negligible, which can be obtained from a series of possible formation mechanisms~\cite{DeLuca:2019buf,1704.06573,1901.05963}
(for the formation of PBHs with near-extremal spin, see e.g.~\cite{Harada:2016mhb,Harada:2017fjm,Cotner:2017tir}).
The emission rate of particle species $i$ per unit time $t$ and total energy
$E$ from Hawking radiation is given by~\cite{Hawking:1975iha}
\begin{equation}
\frac{\text{d}^{2}N_{i}}{\text{d}t\text{d}E}=\frac{g_{i}\,\Gamma_{i}}{2\pi}\left[\text{exp}\left(\frac{E}{T_{\text{PBH}}}\right)-(-1)^{2s_{i}}\right]^{-1}\,,\label{eq:HR_spectrum}
\end{equation}
where $s_{i}$ and $g_{i}$ are the spin and the total degree of freedom
of the particle $i$, respectively, and  the temperature $T_{\text{PBH}}$
of a PBH with mass $M_{\text{PBH}}$ is~\cite{Page:1976df} 
\begin{equation}
T_{\text{PBH}}\approx10.6\times \left(\frac{10^{15}\,\text{g}}{M_{\text{PBH}}}\right)\,\text{MeV}\,.\label{eq:BH_temperature}
\end{equation}
The graybody factor $\Gamma_{i}$ in Eq.~\eqref{eq:HR_spectrum} is
determined by the equation of motion of the particle in curved space
time near the horizon. It describes the probability that the particle
$i$, created at the PBH horizon, finally escapes to spatial infinity.
In the geometric optics limit (i.e., the high-energy limit), the graybody
factor for electron can be approximated as $\Gamma_{e}\simeq27\,G^{2}\,M_{\text{PBH}}^{2}\,E^{2}$.
Note that Eq.~\eqref{eq:HR_spectrum} only describes the primary particles
directly emitted from PBHs. The decays of the unstable primary particles
can produce secondary stable particles. In the calculation of the
graybody factor and the energy spectra of the emitted particles, we use
the numerical code of ${\tt {BlackHawk}}$~\cite{Arbey:2019mbc,Auffinger:2022sqj} in which both the primary and secondary production processes are calculated. 
For the low-energy particle production and decay, we use the results from the \texttt{Hazma}
code \cite{1907.11846} which is now included in the updated version {\tt {BlackHawk-v2.1}}~\cite{Auffinger:2022sqj}
.  

\begin{figure}
\marginnote{\scriptsize 1.3}
\centering \includegraphics[width=16cm]{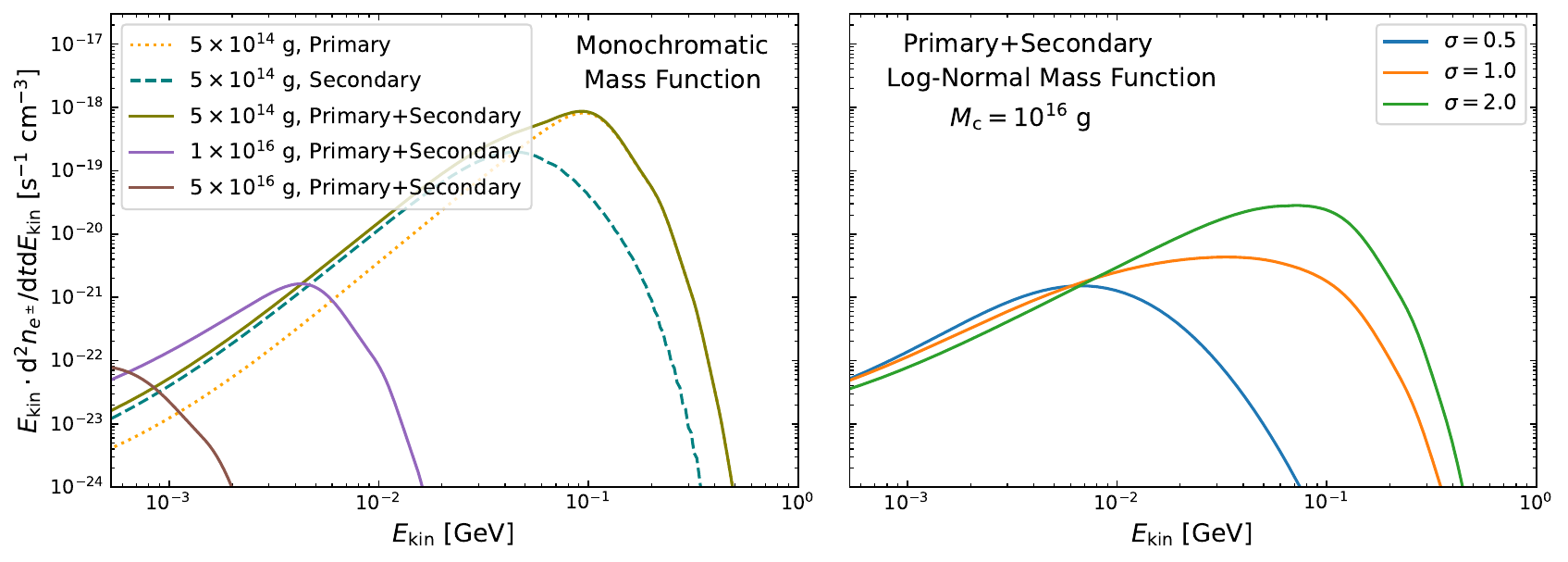} 
\caption{Left) Initial fluxes of CR positron evaporated from PBHs with monochromatic mass distribution with $M_\text{c}=5\times10^{14}$~g, $ 1\times 10^{16}$~g, and  $5\times 10^{16}$~g and energy density $\rho_\text{PBH} = 0.43\,\text{GeV}\,\text{cm}^{-3}$. For the case of light PBHs with  $M_\text{c}=5\times10^{14}$~g, the contribution from the secondary positrons are also shown as they are non-negligible.
Right) The same as the left but for log-normal mass distribution with $M_\text{c}=10^{16}$~g and three different widths $\sigma=0.5$, 1.0 and 2.0. }
\label{fig:secondary_components}
\end{figure}

The energy spectrum $\text{d}^{2}n_{e^{\pm}}/\text{d}t\,\text{d}E_{\text{kin}}$ from all
the evaporating PBHs with different masses is given by
\begin{equation}
\frac{\text{d}^{2}n_{e^{\pm}}}{\text{d}t\,\text{d}E_{\text{kin}}}=\int_{M_{\text{min}}}^{\infty}\text{d}M_{\text{PBH}}\frac{\text{d}^{2}N_{e^{\pm}}}{\text{d}t\,\text{d}E_{\text{kin}}}\frac{\text{d}n}{\text{d}M_{\text{PBH}}}\,,\label{eq:emission_rate}
\end{equation}
where $\text{d}n/\text{d}M_{\text{PBH}}$ is the mass distribution
function of PBH. Depending on formation mechanisms, the mass function
can be a peak theory distribution~\cite{Tashiro:2008sf,Germani:2018jgr}, log-normal distribution~\cite{Kannike:2017bxn}, monochromatic distribution~\cite{Carr:2017jsz} or power-law distribution~\cite{Carr:2017edp}.
In this work, we consider two widely used models, namely, monochromatic
and log-normal mass functions. A nearly monochromatic mass function
is naturally expected if all PBHs are formed at a same epoch, and
a log-normal mass function can arise from inflationary fluctuations~\cite{Dolgov:1992pu,Clesse:2015wea}. The two mass functions are given by 
\begin{equation}\label{eq:mass_func}
\frac{\mathrm{d}n}{\mathrm{d}M_{\text{PBH}}}=
\left\{ \begin{aligned} & A_1\,\delta(M_{\text{PBH}}-M_{\text{c}}) &\text{(monochromatic)}
\\
 & \frac{A_2}{\sqrt{2\pi}\sigma M_{\text{PBH}}^{2}}\,\text{exp}\bigg[-\frac{\text{ln}^{2}(M_{\text{PBH}}/M_{\text{c}})}{2\sigma^{2}}\bigg]
 &\text{(log-normal)},
\end{aligned}
\right.
\end{equation}
where 
$M_{\text{c}}$ is the characteristic mass,
$\sigma$ is the width of the log-normal mass distribution,
$A_1$ and $A_2$ are normalized factors
with different unit, which are determined by $\rho_\text{PBH}$ the energy density of PBH
as follows
\begin{equation}\label{eq:massfunction}
\int_{M_{\text{min}}}^{\infty}M_{\text{PBH}}\frac{\mathrm{d}n}{\mathrm{d}M_{\text{PBH}}}\mathrm{d}M_{\text{PBH}}=\rho_{\text{PBH}} .
\end{equation}
Due to Hawking radiation, PBHs continually lose their masses. Analytical and numerical calculations both confirmed that the lighter the PBH, the faster the evaporation~\cite{MacGibbon:1990zk,Chao:2021orr},
suggesting PBHs with very small masses $(\lesssim5\times10^{14}\,\mathrm{g})$
are absent by now. 
In this work we only take into account the contribution from  the
existing PBHs, the lower bound $M_{\mathrm{min}}$ in Eq.~\eqref{eq:massfunction}
is fixed at $M_{\text{min}}=5\times10^{14}\,\mathrm{g}$. From Eq.~\eqref{eq:massfunction}, the relation between the normalized factor and $\rho_\mathrm{PBH}$ can be obtained. 
It is evident that $A_1 = \rho_\mathrm{PBH}/M_\mathrm{c}$ for $M_\mathrm{c} > M_\mathrm{min}$.  
The value of $A_2$ can be expressed as
$A_2 = \rho_\mathrm{PBH}/k(\sigma)$, 
where 
$k(\sigma) \equiv \int_{M_\text{min}}^\infty\frac{ \text{d} M_\text{PBH}  }{\sqrt{2\pi}\sigma M_{\text{PBH}}} \text{exp}[-\frac{\text{ln}^{2}(M_{\text{PBH}}/M_{\text{c}})}{2\sigma^{2}}]$. 
For typical $\sigma$ in the range 0.5-2.0, the value of $k$ is found to be in the range $0.5\lesssim k \lesssim 1$ from numerical calculations.
The case of $k=1$ corresponds to  the limit of $M_\mathrm{min} = 0$.

In the left panel of Fig.~\ref{fig:secondary_components}, 
we show the calculated emission spectrum 
$\text{d}^{2}N_{e^{\pm}}/\text{d}t\,\text{d}E_\text{kin}$ for CR $e^{\pm}$ as a function of kinetic energy $E_{\text{kin}}=E-m_{e}$ in the monochromatically distributed PBHs with typical masses
$M_{\text{c}}=5\times10^{14}\ \text{g}$, $1\times10^{16}\ \text{g}$ and 
$5\times10^{16}\ \text{g}$, respectively. 
For light PBHs with  $M_{\text{c}}=5\times10^{14}\ \text{g}$, the contributions from secondary particles are important, the secondaries can be dominant in the low-energy region below 10 MeV and can change the spectral shape significantly.
For heavier PBHs with   $M_{\text{c}}\gtrsim 1\times10^{16}\ \text{g}$, the secondary contribution is negligible, so only the total contribution is shown.
In the right panel of Fig.~\ref{fig:secondary_components}, 
the energy spectra for the case of log-normal PBH mass distribution is shown for fixed $M_\text{c}= 1\times10^{16}\ \text{g}$ with three different widths of $\sigma=0.5, 1.0$ and $2.0$.
Compared with the left panel, 
for larger width such as $\sigma=2.0$, the spectra can extend to much higher energies, since the number density of lighter PBHs is significantly increase with given energy density $\rho_\text{PBH}$.

\section{CR propagation in the Galaxy and the heliosphere }\label{sec:propagation}

\subsection{CR propagation in the Galaxy} 

We use the two-dimensional (2D) diffusion models for CR propagation in the Galaxy. In our models,  the diffusion zone is assumed to be a cylinder with radius 
$R_h\approx 20$ kpc and half-height $z_h = 1 \sim 10$~kpc~\cite{Berezinsky:1990qxi,Webber:1992dks}.
The diffusion equation of
CR charged particles can be written as~\cite{Berezinsky:1990qxi,Strong:2007nh}
\begin{align} \label{eq:propagation}
	\frac{\partial \psi}{\partial t} =&
	q(\mathbf{r},p)+\nabla\cdot(D_{xx}\nabla \psi-\mathbf{V}_c  \psi)+
	\frac{\partial}{\partial p}p^2 D_{pp}\frac{\partial}{\partial p}\frac{\psi}{p^2}
	\nonumber\\ 
	& - \frac{\partial}{\partial p} 
	\left[ 
	\dot{p}\psi -\frac{p}{3}(\nabla\cdot\mathbf{V}_c) \psi  \right] -\frac{\psi}{\tau_f}  -
	\frac{\psi}{\tau_r} ~, \end{align} %
where 
$ \psi(\mathbf{r},p,t)$ is the number density per unit of particle momentum $p$ at
the position $\mathbf{r}$ which is related to the phase space distribution function $f(\mathbf{r},\mathbf{p},t)$ as $\psi(\mathbf{r},p,t) =4\pi p^2 f(\mathbf{r}, \mathbf{p},t)$, 
$q(\mathbf{r},p)$ is the time-independent source term, 
$D_{xx}$ is the energy-dependent spatial diffusion coefficient, 
$\mathbf{V}_c$ is  the convection velocity  related to the galactic wind, 
$D_{pp}$ is the diffusion coefficient in momentum space, 
$\dot{p} \equiv \text{d}p/\text{d}t$ is the momentum loss rate,  $\tau_f$ and $\tau_r$ are
the time scales of particle fragmentation and radioactive decay, respectively.
%
%
For the boundary conditions, it is assumed that particles can escape freely at the boundary of the diffusion halo. In  cylinder coordinates, the boundary condition  corresponds to $ \psi(R=R_h,z,p)=\psi(R, z=\pm z_h, p)= 0$, where $R$ and $z$ are the cylinder radius and  the height, respectively. For stable CR sources, the steady-state solution can be achieved, the corresponding condition is $\partial \psi/\partial t = 0$. 
The diffusion equation can be numerically solved use   public codes, such as the  \texttt{Galprop} code \cite{Strong:1998pw,Moskalenko:2001ya,Strong:2001fu,Moskalenko:2002yx,Ptuskin:2005ax}.

In the diffusion equation, the diffusion coefficient $D_{xx}$ represents the CR scattering with the random  magnetic fields, which can be  parameterized as follows
\begin{align}
	D_{xx} = \beta^\eta D_0 \left(\frac{\rho}{\rho_0}\right)^{\delta} ,
\end{align}
where $\rho = p/(Ze)$ is the rigidity of the CR particle with electric charge $Ze$, $D_0$ is a normalization constant determined at a reference rigidity $\sim4$~GV, $\delta$ is the spectral power index, for Kolmogorov type of turbulence $\delta =1/3$~\cite{1941DoSSR..30..301K}, and for an Iroshnikov-Kraichnan cascade $\delta =1/2$~\cite{1964SvA.....7..566I,10.1063/1.1761412},
$\beta = v/c$ is the velocity of CR particles relative to the speed of light $c$, and    $\eta$ is a parameter introduced to accommodate the low-rigidity behavior of the CR spectra.  
If necessary, {\it ad hoc} breaks in the power-law behavior in rigidity can be introduced. For instance, in one-break case, $\delta$ can take two different values of $\delta_{0(1)}$ for $\rho$  below (above) a reference rigidity $\rho_0$. For multiple breaks, $\delta$ can take different values of $\delta_i$ in the rigidity region  $\rho_{i-1} \leq \rho \leq \rho_{i}$, where $\rho_i$ are the corresponding reference rigidities.

The diffusion in momentum space  is described by the parameter $D_{pp}$ which can be parameterized as follows~\cite{Berezinsky:1990qxi,1994ApJ...431..705S}
\begin{align}\label{eq:Dpp}	
	D_{pp} = \frac{4 V_a^2\,p^2}{3 D_{xx}\, \delta\,(4-\delta^2)(4-\delta)},
\end{align}
where $V_a$ is the Alfv\`{e}n velocity which characterizes the 
propagation of disturbances in Galactic magnetic fields.
The scattering of charged particles by the random motion of the magnetic fields characterized by the  Alfv\`{e}n velocity leads to a certain amount of second-order Fermi acceleration during propagation, which can significantly modify the low-energy CR spectra.

In this work the value of $V_a$ is considered to be constant in the whole diffusion halo, which should be understood as an effective parameter.  In some semi-analytic framework, the  re-acceleration is  assumed to be confined in the galactic disk normalized to a half-width of $h\sim 0.1$~kpc, which allows for fast analytical calculations for CR propagation (see e.g.~\cite{Maurin:2002ua}). The value of $V_a$ obtained in the semi-analytical approach should be roughly rescaled by a factor of $\sqrt{h/z_h}$ when compared with the one adopted in this work.

The convection velocity $\mathbf{V}_{c}$ is modeled as a vector field perpendicular to the galactic disk, starting from $V_{c0}$ at  $z=0$ and increases linearly with $z$ with gradient $\text{d}V_{c}/\text{d}z$. 
The expression is 
$\mathbf{V}_{c} =\hat{\mathbf{e}}_z (V_{c0}+(\text{d}V_c/\text{d}z)\cdot |z|) \cdot z/|\mathbf{z}|$,
where $\hat{\mathbf{e}}_z$ is the unit vector alongside the direction of the $z$ axis.

CR electrons/positrons loss energy through the process of ionization, Coulomb scattering, bremsstrahlung in the neutral and ionized medium, and also through synchrotron radiation and inverse Compton scattering (ICS). For the kinetic energy below GeV, the energy loss is dominated by ionization, Coulomb scattering. Above GeV, the energy loss is dominated by synchrotron radiation and ICS processes
~\cite{Moskalenko:1997gh,Strong:1998pw}.
%
The atomic hydrogen (HI) distribution is represented by 
$n_\text{HI}(R,z)=n_\text{HI}(R)\exp[-2\ln 2(z/z_0)^2]$,
where $n_\text{HI}(R)$ and $z_0$ are taken from~\cite{1976ApJ...208..346G,1986A&A...155..380C}.
The distribution of molecular hydrogen ($\text{H}_2$) is taken as
$n_{\text{H}_2}(R,z)=n_{\text{H}_2}(R)\exp[-2\ln 2(z/70\text{pc})^2]$
from CO surveys and  $n_{\text{H}_2}(R)$ is from~\cite{1988ApJ...324..248B}.
The distribution of the ionized gas (HII) is taken as the two-component model from 
~\cite{1991Natur.354..121C}.
The interstellar radiation field (ISRF) is relevant for the leptonic energy losses via ICS process. The low-energy photons involved in this process originate from stars, and are further reprocessed by Galactic dust; CMB photons also contribute with a comparable energy density.
In the {\tt Galprop} code, the  ISRF calculation uses emissivities based on stellar populations and dust emission. The infrared emissivities per atom of HI, H$_2$ are based on COBE/DIRBE data from~\cite{1997ApJ...480..173S}, combined with the distribution of HI and $\text{H}_2$. 

The primary CR electrons are believed to be accelerated by supernova
remnants (SNR) and pulsar wind nebulae. 
The distribution of the primary CR  sources can be modeled by a spatial density
function multiplied by a broken power-law spectrum with $m$-fold breaks at rigidities $\rho_i, (i=1,\dots,m)$ with indices $\gamma_i$ before each break
\begin{equation}\label{eq:src-pri-1}
q(\mathbf{r},p)=n(\mathbf{r})
\left(\frac{\rho}{\rho_{0}}\right)^{-\gamma_{0}}
\prod_{i=0}^{m-1}\left[\frac{\max(\rho,\rho_{i})}{\rho_{i}}\right]^{\gamma_{i}-\gamma_{i+1}}\,.
\end{equation}
For instance, if only one break at $\rho_0$ is considered, the source power term has two power indices $\gamma_0$ and  $\gamma_1$.
The spatial distribution of the primary sources is assumed to follow  that of SNRs, which is parameterized as follows
\begin{equation}\label{eq:src-dist-pri-1}
n(R,z)\propto
\left(\frac{R}{r_{\odot}}\right)^{a}
\exp\left(-b\frac{R-r_{\odot}}{r_{\odot}}\right)\exp\left(-\frac{|z|}{0.2~\text{kpc}}\right),
\end{equation}
where the two source parameters $a$ and $b$ slightly depends on CR species. 

%

In addition to the primary CR electrons,
there are secondary CR $e^\pm$  created by
the collision of primary CR nuclei with the interstellar medium (ISM). 
The secondary source term for CR electrons and positrons is given by
%
\begin{equation}\label{eq:src-sec}
	q_{e^{\pm}}^{\text{sec}}(\mathbf{r},p)
	=
	\sum_{ij}n_{j}(\mathbf{r})\int\mathrm{d}p_{i}~c\,\beta_{i}~\psi_{i}(\mathbf{r},p_{i})~\frac{\mathrm{d}\sigma^{ij\rightarrow e^{\pm}}(p,p_{i})}{\mathrm{d}p},
\end{equation}
where 
the index $i$ runs through primary CR particles such as proton and Helium,
$\beta_{i}$ is the velocity of primary CR particle $i$,
$n_{j}$ is the number density of the $j$-th ISM component with
$j$ runs through HI, HII, and H$_{2}$,
$d\sigma^{ij\rightarrow e^{\pm}}(p,p_{i})/\mathrm{d}p$ is the differential cross-section for creating a secondary  $e^\pm$ with momentum $p$ from an incident primary particle $i$ with momentum $p_i$. Secondary $e^\pm$ are typically produced from the decay of $\pi^\pm$ and $K^\pm$ during the collision with ISM~\cite{Moskalenko:1997gh}.

\subsection{CR propagation in the Heliosphere}
In the vicinity of the Sun, the propagation of charged CR particles is affected
by the regular and irregular heliospheric magnetic fields generated by the out
flowing solar wind. The solar activity leads to time-dependent suppression
of the CR particle flux with rigidity below $\sim50$~GV, which is referred
to as the solar modulation effect. 
%
In this work we consider two approaches for estimating the solar modulation effect. One is based on numerical solution of the Parker equation, the other one is based on the force-field approximation. 

\subsubsection{The Parker equation}
The CR propagation in the heliosphere can be described by the following Parker equation
\begin{align}\label{eq:Parker_eq}
\text{\ensuremath{\frac{\partial U}{\partial t}=\nabla\cdot(K^{S}\nabla U-(\mathbf{V}_{\text{sw}}+v_{d})U)+\frac{1}{3}}\ensuremath{\nabla}$\cdot$
\ensuremath{\ensuremath{\mathbf{V}_\text{sw}\frac{\partial}{\partial T}\left(\alpha_{\text{rel}} E_\text{kin} U\right)}}},
\end{align}
where $U(\mathbf{r},t,E_\text{kin})$ is the particle number density per kinetic energy $E_\text{kin}$,
$K^{S}$ is the symmetric part of the diffusion tensor, 
$V_{\text{sw}}$ is the solar wind velocity, 
$v_{d}$ is the particle magnetic drift velocity, and $\alpha_{\text{rel}}=(E_\text{kin}+2m)/(E_\text{kin}+m)$. 
The component of $K^S$ parallel to the magnetic field $K_{||}$ is believed to be dominant over that perpendicular component $K_{\perp}$.  The parallel component is
parametrized as
\begin{align}
	K_{||}=\frac{\beta}{3}K_0 \left(\frac{\rho}{1~\text{GV}+g_\text{low}}\right)\left( 1+\frac{r}{1~\text{AU}}\right),
\end{align}
where 
$K_0$ is the diffusion parameter and $r$ is the distance to the Sun, 
$g_\text{low}$ is a parameter depends on the solar activity.
The $i$-th perpendicular component $K_{\perp,i}$ is related to the  parallel component through 
$K_{\perp,i}=\rho_i \,K_{||}$.  The parameters $g_\text{low}$ and $\rho_i$ are two major free parameters which need to be determined by the CR data measured at different periods of solar activity.
The Parker equation can be numerically solved using the code \texttt{Helmod}~\cite{Bobik:2011ig,Bobik:2016,Boschini:2017gic,Boschini:2019ubh,Boschini:2022jwz}
in which the propagation parameters are calibrated using the up-to-date CR data. 
In the \texttt{Helmod} code, the CR flux at TOA  is related to that in the LIS as follows
\begin{align}\label{eq:helmod_sol}
    \Phi^{\text{TOA}}(\rho)=\int \Phi(\rho)G(\rho,\rho') d\rho' ,
\end{align}
where $G(\rho,\rho')$ is the normalized probability for a CR particle with a rigidity $\rho'$ in the LIS but 
observed at TOA with a rigidity $\rho$. The function $G(\rho,\rho')$ can be obtained from the \texttt{Helmod} python module which provides the numerical values with rigidity binning and time interval according to the given CR measurements.
The \texttt{Helmod}  code is able to  quantitatively reproduce the time variation of CR fluxes such as that of protons,  and the predicted LIS proton flux is in remarkable agreement with the Voyager-1 data~\cite{Boschini:2019ubh}. 
Other numerical codes solving the Parker equation include {\tt SOLARPROP}~\cite{Kappl:2015hxv}
and {\tt HELIOPROP}~\cite{Vittino:2017fuh}, etc.
%

\subsubsection{Force field approximation}
Another commonly-adopted   model for solar modulation is based on the simplified force-field
approximation~\cite{Gleeson:1968zza}. In this approach, the CR flux $\Phi$ at TOA is related to that at LIS through the relation 
\begin{align}\label{force-field}
\Phi^{\mathrm{TOA}}(E^\text{TOA}_\text{kin})=
\left(\frac{2m E^\text{TOA}_\text{kin} +(E^\text{TOA}_\text{kin})^2}{2m E_\text{kin}+E_\text{kin}^2}\right)
\Phi(E_\text{kin}) .
\end{align}
The kinetic energy $E_\text{kin}$ of the CR particle in LIS is related to that at TOA
through the relation $E_\text{kin}=E_\text{kin}^\text{TOA}+e\,\phi_F\,|Z|$, where $\phi_F$ is the so-called Fisk potential which is a phenomenological parameter that needs to be determined together with  other propagation parameters.  In this method, the solar modulation effect is assumed to be homogeneous. 
The value of $\phi_F$ is dependent on the CR species, and different values of $\phi_F$ should be adopted to fit the same CR specie measured at different time period. There is a strong degeneracy between $\phi_F$ and the CR propagation parameters and the primary source parameters.

\subsection{Benchmark propagation models}\label{subsec:benchmark_propagation_models}
%
As shown in Eq.~\eqref{eq:Dpp}, re-acceleration is generically linked to the spatial diffusion of CR particles, and has important phenomenological consequences. It is known that re-acceleration can provide a natural mechanism to reproduce the low-energy B/C ratio with Kolmogorov type of turbulence and a spectral break in the power-law spectrum of primary CRs at a rigidity of few~GV~\cite{Simon:1996dk}.
Currently, diffusive re-acceleration models with a sizable $V_a\sim 30$~km/s has received strong support from a number of independent analyses~\cite{Trotta:2010mx,Jin:2014ica,Yuan:2018lmc,Boschini:2017fxq,Boschini:2019gow,Boschini:2020jty,DeLaTorreLuque:2021yfq,Luque:2021nxb}.
For instance, a global fit to the B/C and Carbon data from  HEAO-3, ACE and ATIC-2 using the {\tt Galprop} with force-field approximation for solar modulation found $V_a=38.4\pm 2.1~\text{km/s}$~\cite{Trotta:2010mx}. 
Similar updated analysis using the AMS-02 proton and B/C data gave $V_a=44.6\pm 1.2~\text{km/s}$~\cite{Jin:2014ica}.
%
A recent analysis including the AMS-02 data of  Be, B, C, N, O nuclei also favored a large $V_a= 30\pm2.5~\text{km/s}$~\cite{Johannesson:2016rlh}.  
It was also noticed that fit to another data set of proton, antiproton and Helium tended to give a smaller value of $V_a$, which suggest that different CR species may explore different parts of the ISM~\cite{Johannesson:2016rlh}. 
%
%
In an analyses used the {\tt Helmod} code for calculating the solar modulation,  fitting to the  AMS-02 B/C, proton and He data gave $V_a=28.6\pm3~\text{km/s}$ \cite{Boschini:2017fxq}. 
The result was confirmed by the updated analyses including more secondaries such as Be and Li~\cite{Boschini:2019gow,Boschini:2020jty}.
For the analyses using the numerical code {\tt Dragon} for CR propagation, a recent analysis considered a set of cross sections for the production of secondaries~\cite{DeLaTorreLuque:2021yfq} which are different from those adopted by the {\tt Galprop} code, and found $V_a\sim 30~\text{km/s}$ from fitting to the AMS-02 data of B/C, B/O, Li/C, Li/O flux ratios~\cite{Luque:2021nxb}.
%

Since the effects of different propagation parameters and the injection spectrum are partially degenerated, it is possible to construct alternative propagation models without re-acceleration but include more complicated energy dependence in the diffusion term. A known example is to introducing an additional break in rigidity-dependent diffusion coefficient $D_{xx}$ at rigidity around 4~GV and assuming $V_a=0$. This type of diffusion break (DB) models can reproduce the similar structure in B/C flux ratio. So far both scenario can well explain the data of secondary/primary flux ratio~\cite{Korsmeier:2021brc,Genolini:2019ewc,Weinrich:2020cmw}. 
Note that it is possible to distinguish the two scenarios in future by considering additional observables. For instance, the existence of $V_a$ can be probed by synchrotron radiations~\cite{Orlando:2013ysa} and a break in the injection primary source can be examined by gamma-ray emission from gas clouds~\cite{Neronov:2011wi}.
%
%
%
%

In this work, we consider a number of \texttt{Galprop}-based propagation models with different methods for solar modulation calculations. We focus on the models with constant $V_a$. In the generic case, $V_a$ is expected to be spatial dependent, as $V_a\propto B/\sqrt{n_{\text{ISM}}}$ where $B$ is the interstellar magnetic field and $n_{\text{ISM}}$ is the number density of the ISM atoms. Since we are focused on CR positrons which cannot propagate to a long distance due to fast energy losses, we assume $V_a$ to be a effective constant in the whole propagation halo.
\begin{itemize}
\item {\bf GH-model}. 
In this model, we take the analysis framework of \texttt{Galprop+Helmod} (GH) where the two numerical codes \texttt{Galprop} and \texttt{Helmod} are combined together to provide a single framework to calculate CR fluxes at different modulation levels and at both polarities of the solar magnetic field~\cite{Boschini:2017fxq,Boschini:2018zdv,Boschini:2019gow,Boschini:2020jty}. This is achieved by an iterative optimization procedure to tune the parameters in both \texttt{Galprop} and \texttt{Helmod} to best reproduce the data set of CR
proton flux measured by PAMELA, BESS and AMS-02. The predicted LIS proton spectrum is in a remarkable agreement with the Voyager-1 data~\cite{Boschini:2019ubh}.
We adopt the parameters determined in~\cite{Boschini:2020jty} which is obtained by a Markov Chain Monte Carlo (MCMC) scan of the parameter space to fit the AMS-02 data of light nuclei. In this model, the best fit halo height is fixed at $z_h = 4$~kpc, 
and the best-fit gradient of the convection velocity is $\text{d}V_{c}/\text{d}z = 9.8\,\text{km}\,\text{s}^{-1}\,\text{kpc}^{-1}$ with $V_{c0} = 0$. The diffusion coefficient
 follows a single power law, i.e., $\delta_0 = \delta_1 = 0.415$.
The injection primary spectra have two low-rigidity breaks located at $\sim 1$~GV and $\sim 7$~GV, respectively, which depend on CR species. For protons, $\gamma_{0(1)}$=2.24 (1.70), $\rho_{0(1)}$=0.95 (6.97), and  for $_6\text{C}$,  $\gamma_{0(1)}$=1.00 (1.10), $\rho_{0(1)}$=1.98 (6.34). The details on the primary spectra for different species can be found in Table 2 of  Ref.~\cite{Boschini:2020jty}.
In this model, the source distribution parameters in  Eq.~\eqref{eq:src-dist-pri-1} are fixed at $a=1.9$ and $b=5.0$, respectively.

\item {\bf MIN-, MED-, MAX- and MAX2-models}.
In this type of models, the simplified force-field approximation of Eq.~\eqref{force-field} for solar modulation is adopted, which allows for fast analytical calculations of TOA fluxes and more flexibility to adjust model parameters in data fitting processes. In order to investigate the uncertainties in $z_h$ due to the well-known $z_h-D_0$ degeneracy, we first consider three propagation models from our previous analysis to the AMS-02 proton and B/C data~\cite{Jin:2014ica}, the so called MIN, MED and MAX models for Galprop with $z_h=$1.8, 3.2 and 6~kpc, respectively. The parameters in the  models are slightly tuned to better reproduce the latest AMS-02 B/C 2021 data~\cite{AMS:2023anq}.
In addition, we also include a model with $z_h=10$~kpc (MAX2 model) which is derived from the MAX model. 
In all these models, the convection effect is not considered 
(i.e., $\text{d}V_c/\text{d}z = 0$ and $V_{c0} = 0$), and  the power index of the diffusion coefficient is close to the Kolmogorov type $\delta_0 = \delta_1 \sim 0.3$.
The Fisk potential is fixed at $\phi_F=550$~MV. 
The CR primary spectra are assumed to have the same power-law index as that of CR protons which as a break at a reference rigidity $\rho_0=4$~GV with $\gamma_{0(1)}=1.79\,(2.45)$ for the MED model. 
In these models, the values of the source  parameters $a=1.25$ and $b=3.56$ are adopted, which are the default values in the  {\tt Galprop} code.
%

\item {\bf DB-model}. 
In this model, the re-acceleration term is switched off, i.e. $V_a=0$, while a break in the diffusion coefficient $D_{xx}$ is introduced at rigidity $\bar{\rho}_0=3.94$~GV with two different diffusion power-law indices $\delta_{0(1)}=-0.98\,(0.49)$. There is also no break assumed in the injection spectrum, i.e., $\gamma_0=\gamma_1=2.4$. The propagation parameters are based on the ``BASE" propagation model proposed in~\cite{Korsmeier:2021brc}. 
In this model, the convection velocity is assumed to be constant $V_{c0} = 3.34\,\text{km/s}$  with $\text{d}V_{c}/\text{d}z = 0$.
The force-field approximation is adopted with the Fisk potential fixed at $\phi_F=610$~MV.
For simplicity, the nuclear cross sections for the secondary production are taken as the default values in the {\tt Galprop} code. Although in this work we focus on the the re-acceleration models, this model serves as a useful counterexample.
\end{itemize}

\begin{table}[th]
\setlength\tabcolsep{11pt}
\centering %
\begin{tabular}{lccccccc}
\hline\hline
Parameters & GH & MIN & MED & MAX & MAX2 & DB & \tabularnewline
\hline 
$z_{h}$ {[}$\kpc${]} & 4.0 & 1.8 & 3.2 & 6.0 & 10.0 & 4.0 & \tabularnewline
$D_{0}$ {[}$10^{28}~\cm^{2}~\text{s}^{-1}${]} & 4.3 & 3.2 & 5.8 & 9.8 & 13.9 & 5.05 & \tabularnewline
$\eta$  & 0.7 & 1 & 1 & 1 & 1 & 1 & \tabularnewline
$\delta_{0}$ & -- & -- & -- & -- & -- & -0.98 & \tabularnewline
$\delta_{1}$ & 0.415 & 0.30 & 0.29 & 0.29 & 0.29 & 0.49 & \tabularnewline
$\bar{\rho}_{0}$ {[}$\GV${]} & -- & -- & -- & -- & -- & 3.94 &\tabularnewline
$V_{a}$ {[}$\km~\text{s}^{-1}${]} & 30 & 38 & 38 & 36 & 36 & -- & \tabularnewline
$\phi_F$ {[}$\GV${]} & -- & 0.55 & 0.55 & 0.55 & 0.55 & 0.61 & \tabularnewline
$\gamma_0$ & 2.24 & 1.75 & 1.79 & 1.81 & 1.81 & 2.36 &\tabularnewline
$\gamma_1$ & 1.70 & 2.44 & 2.45 & 2.46 & 2.46 & -- &\tabularnewline
$\rho_0$ {[}$\GV${]} & 0.95 & 4.0 & 4.0 & 4.0 & 2.5 & -- & \tabularnewline
\hline\hline
\end{tabular}\caption{\label{tab:param-fixed}
Fixed parameters for six CR propagation models GH~\cite{Boschini:2020jty}, MIN, MED, MAX~\cite{Jin:2014ica}, MAX2 and DB. The primary indices $\gamma_{0,1}$ and break rigidity $\rho_0$ are for CR protons. $\bar{\rho}_0$ is the rigidity break in diffusion coefficient $D_{xx}$ in the DB model.}
\end{table}

\begin{figure}[tbp]
	\centering
	\includegraphics[width=16cm]{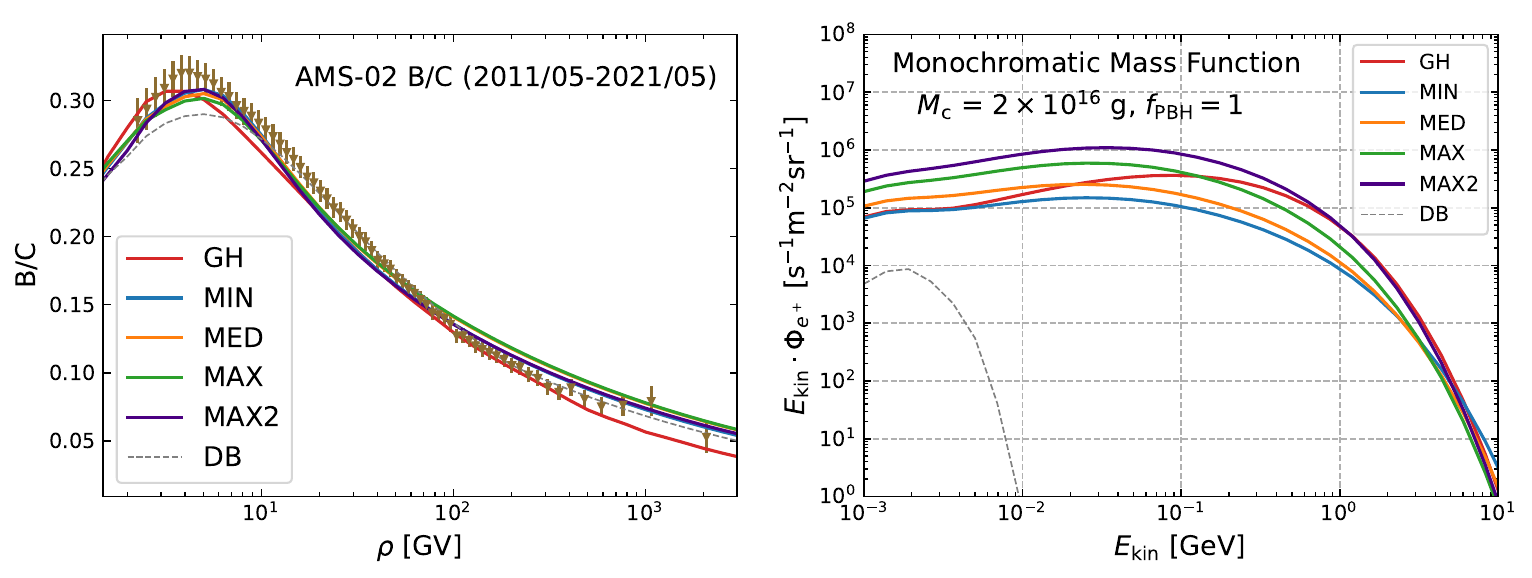}
	\caption{Left) Boron to Carbon flux ratio in the six models listed in Tab.~\ref{tab:param-fixed}. The data measured by the AMS-02 in the period 2011-2021 are shown for comparison. Right) Positron fluxes from PBH evaporation in the six propagation models. The PBH abundance is fixed at $f_\text{PBH}=1$ and mass function is assumed to be monochromatic with $M_\text{c} = 2\times 10^{16}$ g.}
	\label{fig:PBH_fluxes_and_B/C}
\end{figure}

The details of the parameters of these propagation models are listed in Tab.~\ref{tab:param-fixed}. In the left panel of  Fig.~\ref{fig:PBH_fluxes_and_B/C} we show the predicted B/C flux ratio from the six propagation models and compare them with the latest AMS-02 data~\cite{AMS:2023anq}. The figure indicates that although the parameter sets are quite different, all of them can well explain the current data. 

The source term for electrons and positrons from PBHs can be obtained by combining Eq.~\eqref{eq:emission_rate}, Eq.~\eqref{eq:massfunction} and the PBHs spatial distribution in the Galaxy.
For simplicity, the spatial distribution of PBHs is assumed to follow that of DM, namely, 
$\rho_{\text{PBH}}(r)$ satisfies $\rho_{\text{PBH}}(r)=f_{\text{PBH}}\cdot\rho_{\text{DM}}(r)$,
where $\rho_{\text{DM}}(r)$ is the DM density profile. 
In this work, we adopt the Navarro-Frenk-White (NFW)  profile \cite{Navarro:1996gj} 
\begin{equation}
	\rho_\text{DM}(r) = \rho_0\bigg(1+\frac{r_\odot}{r_\text{s}}\bigg)^2\bigg(\frac{r}{r_\odot}\bigg)^{-1}\bigg(1+\frac{r}{r_\text{s}}\bigg)^{-2} \, ,
\end{equation}
where 
$\rho_0 = 0.43\,\text{GeV}\cdot\text{cm}^{-3}$ is the  local DM energy density, 
$r_\odot = 8.5\,\text{kpc}$ is the distance of the Sun to the Galactic center, 
and $r_\text{s} = 20\,\text{kpc}$ is a typical radius parameter.
Note that the dependence of CR electron and positron fluxes on the DM profile is rather weak. 
In the right panel of Fig.~\ref{fig:PBH_fluxes_and_B/C}, 
we show the predicted positron energy spectra for a monochromatic PBH distribution with a typical  mass $M_\text{c}=2\times 10^{16}$~g in the six propagation models. 
It can be seen that after calibrated by the B/C flux ratio,  all the five models with re-acceleration predict similar positron spectra: a significant portion of the evaporated positrons can be boosted to GeV region, which suggest that the AMS-02 positron data can provide useful constraints.
An exception is the DB model, in this model, the positron flux drops rapidly towards higher energies and features a cut off at $\mathcal{O}(\text{MeV})$ as suggested by Eq.~\eqref{eq:BH_temperature}. 
Thus in this case the AMS-02 positron data cannot place useful constraints on $f_\text{PBH}$. 
In this type of models, the constraints can only be obtained from the Voyager all-electron data as previously discussed in~\cite{Stone:2019}.

%

\begin{figure}[tbp]
    \centering
    \includegraphics[width=16cm]{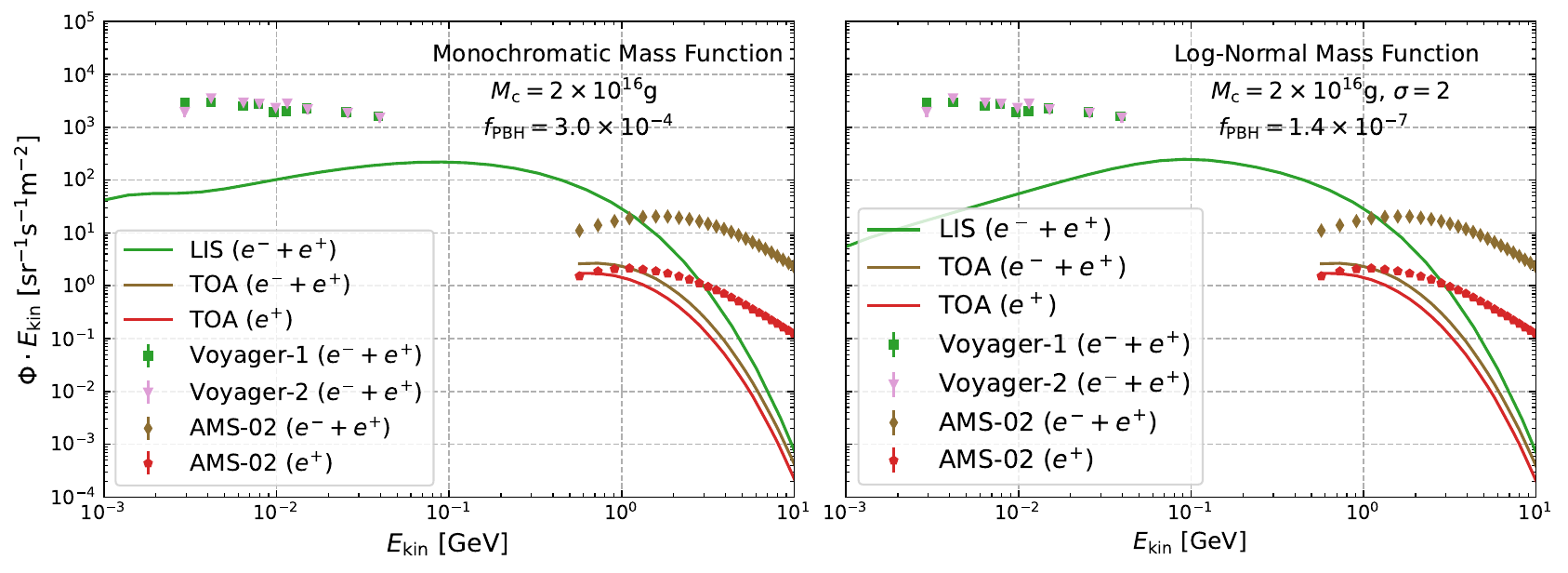}
    \caption{
    Left) LIS and TOA CR positron and all-electron fluxes from the evaporation of PBHs in the GH propagation model. The PBH mass distribution is taken to be monochromatic with $M_\text{c}=2\times 10^{16}$~g. The data of Voyager-1,2 (all-electron)~\cite{Stone:2019} and AMS-02 (positron and all-electron)~\cite{AMS:2021nhj} are also shown. The value of $f_\text{PBH}$ is chosen to saturate the AMS-02 positron data at lowest kinetic energy at 0.5~GeV.
    For the LIS (TOA) fluxes, $E_\text{kin}$ represents the kinetic energy of CR particles at LIS (TOA).
    Right) The same as the left but for PBH log-normal mass distribution with $M_\text{c}=2\times 10^{16}$~g and $\sigma=2.0$.
    }
    \label{fig:2e16_wo_background}
\end{figure}

In the left panel of Fig.~\ref{fig:2e16_wo_background}, 
we show the CR positron and all-electron $(e^-+e^+)$ fluxes predicted from the evaporation of PBHs with monochromatic mass distribution at $M_\text{c}=2\times 10^{16}$~g in the GH propagation model together with the  data of AMS-02 (positron and all-electron) and Voyager-1, -2 (all-electron).
In the figure we only show the predictions of the TOA flux in the same energy region as that of the published AMS-02 data, as the probability function $G(\rho,\rho')$ provided by \texttt{Helmod} is specific for each experiment and data taking period.
The astrophysical background electrons and positrons are not shown for the moment,  such that  conservative constraints (i.e. the constraints without including the astrophysical backgrounds) on $f_\text{PBH}$ can be roughly inferred  from comparing the theoretical predictions with the data.
It can be seen that in the propagation model under consideration, the PBH generated positron flux will be first constrained by the AMS-02 positron data in the GeV region  rather than that from the Voyager all-electron data in the MeV region. 
Since the predicted CR fluxes depend on $f_\text{PBH}$ linearly,  
we find by gradually increasing the value of $f_\text{PBH}$ that for $f_\text{PBH}\approx 3\times 10^{-4}$  the predicted positron flux can saturate the low-energy positron flux at $\sim 0.5$~GeV measured by AMS-02, which  will set the scale of the final constraints from a more robust statistic analysis. 
We find similar results for other propagation models such as the MIN, MED, MAX and MAX2 models.
%
In the right panel of Fig.~\ref{fig:2e16_wo_background}, we show the result for the log-normal PBH mass function with $M_\text{c}=2\times 10^{16}$~g and width $\sigma = 2$ in the propagation model GH. 
The CR backgrounds are also neglected as that in the right panel. 
It can be seen  that in this extended mass function the value of $f_\text{PBH}$ will again be constrained first by the AMS-02 low-energy positron data. A typical constraint is  at the level of $f_\text{PBH} \lesssim 1.4\times 10^{-7}$. 

At present, although the re-acceleration process is generally expected, the exact value of $V_a$ is not yet conclusive.  
Fig.~\ref{fig:2e16_wo_background} suggests that if a large $V_a$ of $\sim 30~\text{km}\cdot\text{s}^{-1}$ is confirmed by the future studies, 
the AMS-02 positron data will be more powerful in constraining $f_\text{PBH}$ than the Voyager and AMS-02  all-electron data.

\section{Constraints on the PBH abundance}\label{sec:constraints}

In this section, we derive the constraints on $f_\text{PBH}$ from the current CR electron and positron data.  

\subsection{Constraints without including astrophysical backgrounds}

In the first step, we derive the constraints under the assumption of null astrophysical backgrounds, which should be conservative and robust.
In calculating the constraints, we adopt a simple approach by requiring 
that the predicted CR flux from PBH evaporation should not exceed 
the experimental data by more than $2~\sigma$ uncertainty in any energy bin, 
which was the approach adopted in Ref.~\cite{Boudaud:2018hqb}, 
and thus allows for a direct comparison with their results.
The constraints are obtained by simply rescaling the value of $f_\text{PBH}$ and compare the predictions with the experimental data.

\begin{figure}[tbp]
    \centering
    \includegraphics[width=13cm]{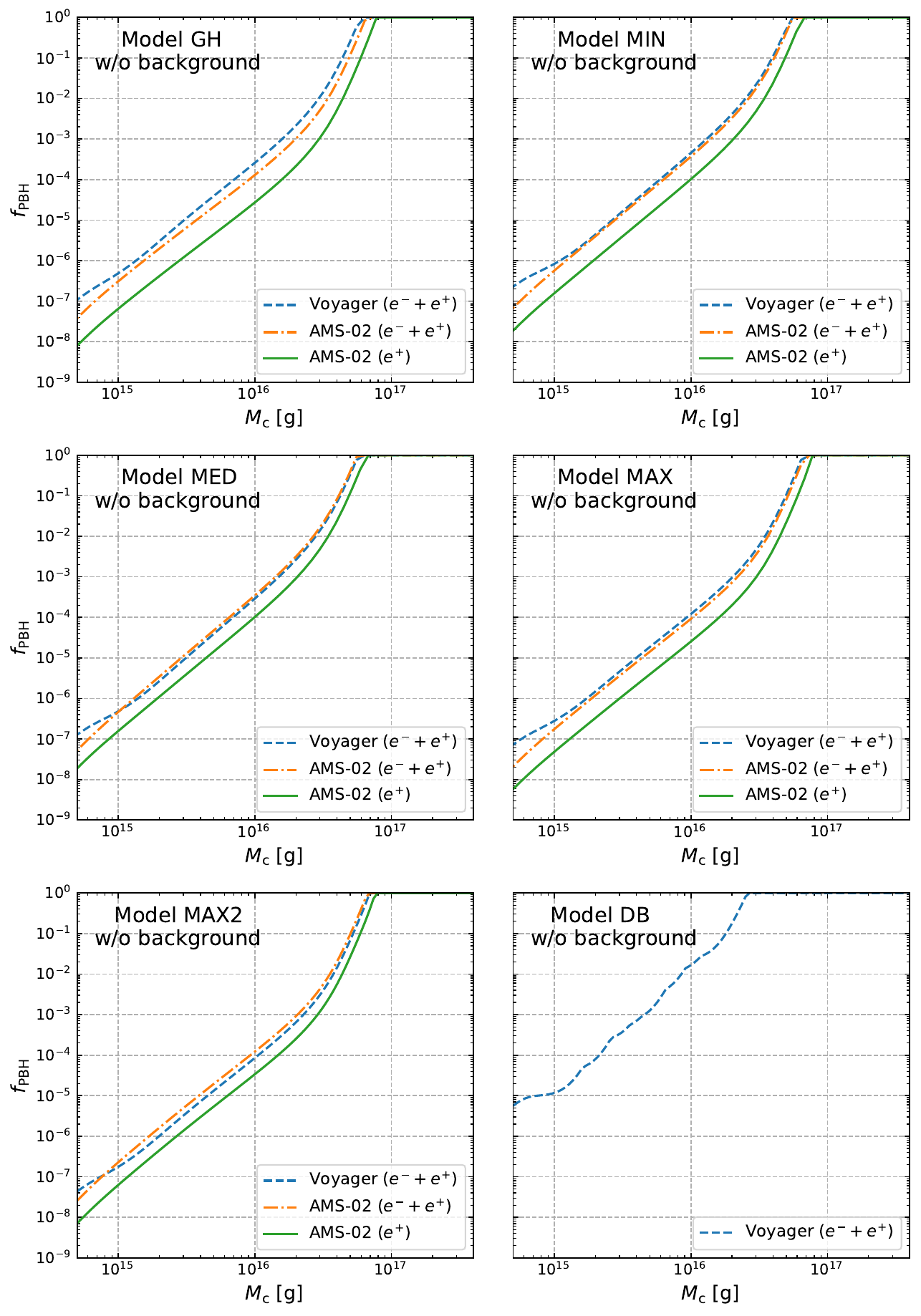}
    \caption{Constraints on $f_\text{PBH}$ in six CR propagation models as functions of the characteristic mass $M_\text{c}$ of the monochromatic mass distribution. The parameters of the propagation models are listed in Tab.~\ref{tab:param-fixed}. The solid-green, dashed-blue, and dot-dashed orange curves represent the constraints obtained individually from the positron data of AMS-02~\cite{AMS:2021nhj}, the all-electron data of Voyager-1,-2~\cite{Stone:2019} and the all-electron data of AMS-02~\cite{AMS:2021nhj}, respectively. }
    \label{fig:exclusion_lines_wo_background}
\end{figure}

In Fig.~\ref{fig:exclusion_lines_wo_background}, 
we show the obtained constraints on $f_\text{PBH}$ in 
the six propagation models from 
the individual data set of Voyager-1, -2 all-electron, AMS-02 all-electron, and AMS-02 positron 
for monochromatically distributed PBHs with characteristic
mass $M_\text{c}$ in the range $5\times 10^{14}- 10^{17}~\text{g}$.
It can be seen that for the GH, MIN, MED, MAX and MAX2 models, 
the constraints from the low-energy AMS-02 positron data are the most stringent. 
As an example,
in the GH propagation model, we obtain conservative upper limit of 
$$
f_{\text{PBH}}\lesssim2.15\times 10^{-4} \ \ \text{at } M_{\text{PBH}} = 2\times 10^{16}~\text{g} ,
$$ 
which is more stringent than that from the Voyager all-electron data by an order of magnitude.
The AMS-02 positron constraints are also more stringent than that from the AMS-02 all-electron data. The reason is that the initial spectra of electrons and positrons from PBH evaporation are the same as the evaporation process is charge symmetric. Thus the prediction for all-electron flux from PBH evaporation should be approximately twice the positron flux. However, the all-electron flux measured by AMS-02 is around an order of magnitude higher than of positron flux. Thus the corresponding constraints are significantly weaker in the null background case.
The constraints derived in Fig.~\ref{fig:exclusion_lines_wo_background} are in a good agreement with the values previously estimated from the left panel of Fig.~\ref{fig:2e16_wo_background}.

Our analysis also show that value of $V_a$ plays an important role in  constraining $f_\text{PBH}$ from the CR data.  As demonstrated in the right panel of Fig.~\ref{fig:PBH_fluxes_and_B/C},  in the case where $V_a$ is vanishing (model DB), the predicted positron flux exhibits a sharp cutoff at kinetic energies $\sim 10\,\text{MeV}$, far below the reach of the AMS-02 experiment.  
Thus for the DB model, only the constraints can only be obtained from the 
Voyager all-electron data, as shown in the lower-right panel of Fig~\ref{fig:exclusion_lines_wo_background}.
%
%

\begin{figure}
    \centering
    \includegraphics[width=15cm]{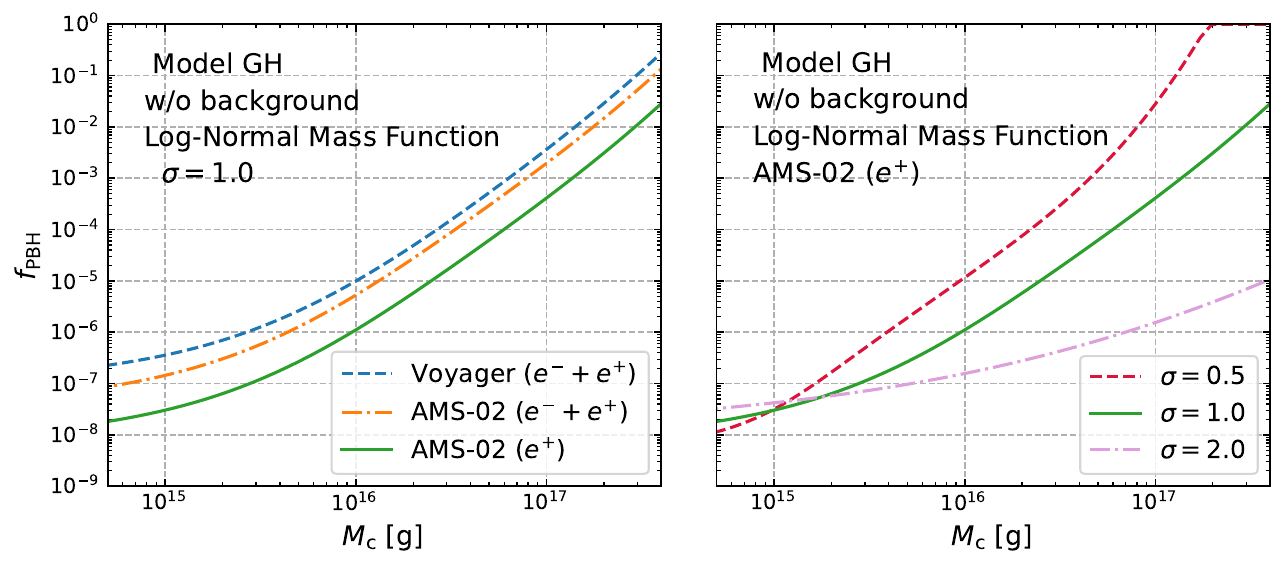}
    \caption{
    Left) Constraints on $f_\text{PBH}$ in the GH propagation model with parameters listed in \tab{tab:param-fixed}. The blue, orange, and green curves represent the constraints obtained individually from the all-electron data of Voyager-1,-2~\cite{Stone:2019}, the all-electron data of AMS-02~\cite{AMS:2021nhj}, and the positron data of AMS-02~\cite{AMS:2021nhj}, respectively. The log-normal PBH mass function with width $\sigma=1.0$ is assumed.
    Right) The constraints from the AMS-02 positron data only, for the log-normal  PBH mass function with different widths of $\sigma=0.5$, 1.0 and 2.0, respectively. }
    \label{fig:excl_log_wo_background}
\end{figure}

In the left panel of  Fig.~\ref{fig:excl_log_wo_background} we show the constraints for log-normally distributed PBHs with a typical width $\sigma=1$ in the GH propagation model. Similar to the case with monochromatic mass function, the AMS-02 positron data provide the most stringent limits in the case with log-normal mass function.
In the right panel of  Fig.~\ref{fig:excl_log_wo_background},  the constraints for different widths $\sigma=0.5, 1.0$ and 2.0 for the log-normal mass function are compared. 
For the log-normal mass function  constraints becomes more stringent as the value of $\sigma$ increases, which is consistent with the flux predictions in the right panel of Fig.~\ref{fig:2e16_wo_background}.

%

\subsection{Constraints with including astrophysical backgrounds}
In the next step, we consider the CR-positron constraints including the secondary positron backgrounds.
Including the astrophysical background in general results in more stringent constraints.
However, so far the theoretical predictions for low-energy (sub-GeV) positron flux still suffer from significant uncertainties. 
In addition to the uncertainties in  propagation models such as the diffusion halo half-height $z_h$, 
large uncertainties arise from  the low-energy $pp$ and $pA$ inelastic scattering cross sections for positron production. 
At present, the cross sections are obtained either from 
analytical parameterizations to the experimental data 
~\cite{Orusa:2022pvp,Kafexhiu:2014cua,Kamae:2006bf}
or from QCD-based Monte-Carlo event generators
~\cite{DelaTorreLuque:2023zyd,Koldobskiy:2021nld,Bierlich:2022pfr}. 
The differences between the approaches are significant.
Other uncertainties specifically related to CR electron and positron energy loss involve that in the  Galactic magnetic fields and  gas distribution (for recently analyses, see e.g.~\cite{DiMauro:2023oqx,DelaTorreLuque:2023zyd}).
Due to these uncertainties,
the derived constraints including the astrophysical backgrounds are 
less robust compared with that without backgrounds.

\begin{figure}[tbp]
    \centering
    \includegraphics[width=15cm]{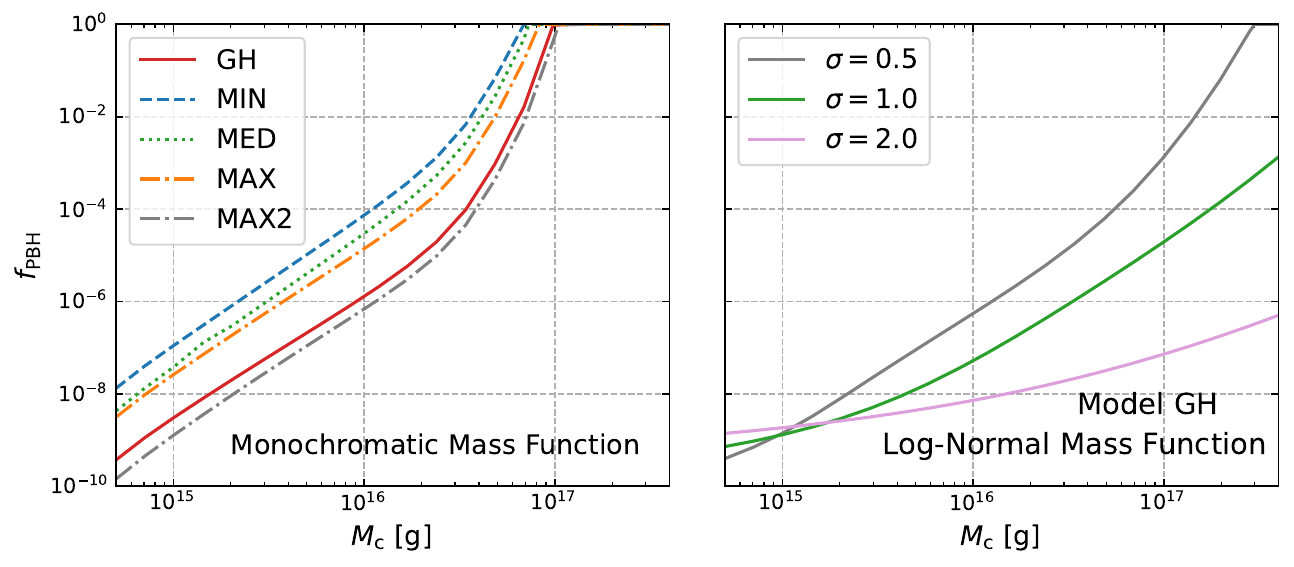}
    \caption{Left) Constraints on $f_\text{PBH}$ from the AMS-02 positron data with the inclusion of astrophysical backgrounds in five CR propagation models. The PBH mass function is assumed to be  monochromatic.
    Right) The constraints for the log-normal PBH mass function with different widths $\sigma = 0.5\,,1.0\,,2.0$ in the GH propagation model.
    }
    \label{fig:excl-w-background}
\end{figure}

To account for the uncertainties in the secondary positron flux, 
we introduce two additional free parameters $N_{e^+}$ and $\delta_{e^+}$ into 
the calculation of the  CR positron flux $\Phi_{e^+}$, namely, 
$\Phi_{e^+}(E_\text{kin})
\to N_{e^+} (E_\text{kin}/\text{GeV})^{\delta_{e^+}}  \Phi_{e^+}(E_\text{kin})$. 
The parameters are determined through fitting to the experimental data. 
%
In the presence of backgrounds, 
we derive the constraints on $f_\text{PBH}$ using 
the standard minimal-$\chi^2$ approach. 
For a given mass function parameters of 
$M_\text{c}$ or a pair of $(M_\text{c}, \sigma)$, 
we first find the best-fit parameters 
$(\hat{f}_\text{PBH}, \hat{N}_{e^+}, \hat{\delta}_{e^+})$ and 
the corresponding  minimal value of 
$\chi^2_\text{min}(\hat{f}_\text{PBH}, \hat{N}_{e^+}, \hat{\delta}_{e^+})$.
Then, we increase the value of  $f_\text{PBH}$ and repeat the fitting process with respect to the other parameters $N_{e^+}, \delta_{e^+}$ and obtain larger values of $\chi^2$. 
This process is repeated until a particular $f_\text{PBH}$ satisfying 
$\Delta \chi^2 = 
\chi^2(f_\text{PBH},\doublehat{N}_{e^+}(f_\text{PBH}),\doublehat{\delta}_{e^+}(f_\text{PBH})) - \chi^2_\text{min}(\hat{f}_\text{PBH}, \hat{N}_{e^+}, \hat{\delta}_{e^+})
= 3.84$ is found.
Here, the parameters with a double hat minimized the $\chi^2$ function for the given $f_\text{PBH}$. 
The obtained values of $f_\text{PBH}$ correspond to the upper limits at the 95\%~C.L.\,.
The AMS-02 positron flux data in the low-energy region $0.5-5.2$~GeV 
(16 data points in total)
are included in the data analysis process.
%
%


%
In the left panel of Fig.~\ref{fig:excl-w-background}, 
we show the obtained constraints on $f_\text{PBH}$ as a function of $M_\text{c}$ in the case of monochromatic mass function in five propagation models. 
The difference between different propagation models are typically within two orders of magnitude.
In the GH model, $\chi^2$ fitting gives $\hat{f}_\text{PBH}=0$, 
$\hat{N}_{e^+}=0.255$, and 
$\hat{\delta}_{e^+} = 0.995$ with 
$\chi^2_\text{min}/\text{d.o.f}=32.51/14$. 
Similar results are found for the MAX2 model. For the model MIN--MAX, the constraints are relatively weaker due to the fact that non-vanishing $\hat{f}_\text{PBH}$ are favored.
The DB model is not considered in the analysis 
as the predicted positron energies from this model are too low to be constrained by the AMS-02 data.
%
In the right panel of Fig.~\ref{fig:excl-w-background}, 
the constraints for the  log-normal mass function with three width $\sigma = 0.5$, 1.0 and 2.0 in the GH propagation model are shown. 
Similar to the right panel of Fig.~\ref{fig:excl_log_wo_background}, increasing $\sigma$ leads to more stringent constraints. 
In all the re-acceleration propagation models, 
we have verified that the constraints from the low-energy AMS-02 positron data are always the most stringent compared with that from the AMS-02 all-electron or the Voyager-1,2 all-electron data.

\begin{figure}
    \centering
    \includegraphics[width=11cm]{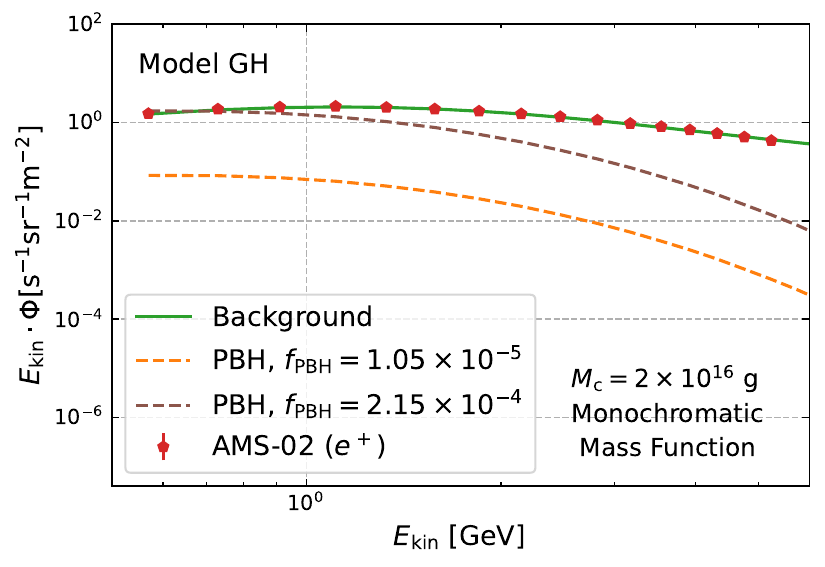}
    \caption{
    Secondary background positron fluxes (solid green) and  
    primary positron fluxes from PBH evaporation (dashed orange)
    corresponding to the derived constraints of 
    $f_\text{PBH}\leq 1.05\times 10^{-5}$ at 95\% C.L. 
    at $M_\text{c} = 2\times 10^{16}\,\text{g}$  (from Fig.~\ref{fig:excl-w-background})
    in the GH model.
    The PBH positron fluxes corresponding to the  constraints of $f_\text{PBH}\leq 2.15\times 10^{-4}$  without including the secondary backgrounds (dashed brown) from \fig{fig:exclusion_lines_wo_background} are also shown 
    for comparison purpose.}
    \label{fig:background+PBH}
\end{figure}

Including the secondary CR positron backgrounds can 
significantly improve the constraints on $f_\text{PBH}$, 
if the model of the secondaries agrees with the data well. 
In Fig.~\ref{fig:background+PBH}, 
we show the calculated secondary positron fluxes correspond to
$M_\text{c} = 2\times 10^{16}\,\text{g}$ and $f_\text{PBH}=1.05\times 10^{-5}$ 
(the value allowed at 95\% C.L. from Fig.~\ref{fig:excl-w-background}) in the GH model.
It can be seen that
the AMS-02 data can be well-fitted by the secondary positron background,
which leads to  more stringent constraints. 
While in the analysis assuming no background, 
the constraints is around  $f_\text{PBH}=2.15\times 10^{-4}$ (according to Fig.~\ref{fig:exclusion_lines_wo_background})
which is around an order of magnitude weaker.
%
%

\begin{figure}[tbp]
    \centering
    \includegraphics[width=11cm]{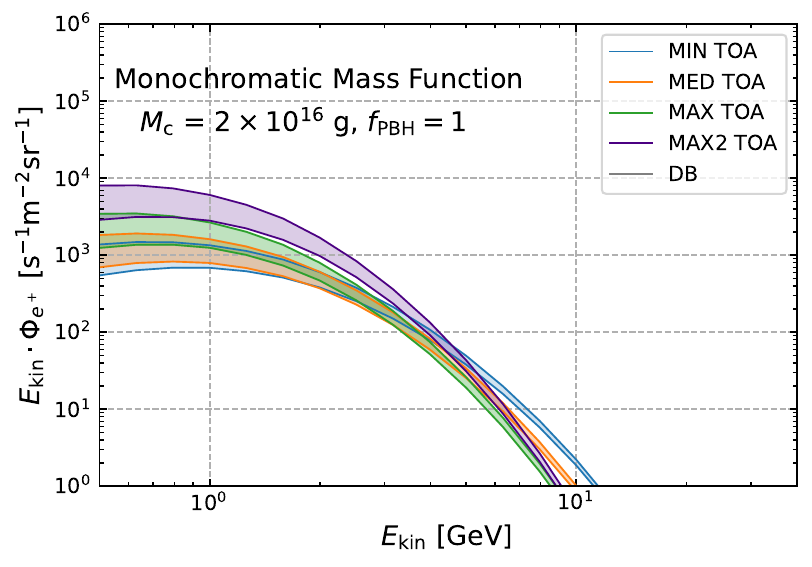}
    \caption{PBH evaporating positron fluxes at TOA in propagation model MIN, MED, MAX and MAX2. 
    The shaded region represents the uncertainty in the positron flux stemmed from the uncertainty of Fisk potential.
    The abundance and mass function of PBHs is consistent with the scenario depicted in the right panel of Fig.~\ref{fig:PBH_fluxes_and_B/C}. }
    \label{fig:force-field-uncertainty}
\end{figure}

The uncertainties in modeling the solar modulation effect is an important source of 
the uncertainties of the obtained constraints, 
as they mainly affect the low-energy CR fluxes at GeV region. 
In order to estimate the uncertainties related to the solar modulation effect, 
we calculate  the CR fluxes in the MIN, MED and MAX models 
with the Fisk potential varying around the default  $\phi_F$ values in Table~\ref{tab:param-fixed} by
$\phi\to \phi_F\pm \Delta\phi_F$, where $\Delta\phi_F$ represents the uncertainty in $\phi_F$ with a typical value of $\Delta\phi=0.1$~GV. The results are shown in Fig.~\ref{fig:force-field-uncertainty}.
It can be seen that the typical uncertainties at $E_\text{kim}\sim 0.5$~GeV 
which is the lowest positron energy measured by the AMS-02 experiment are around a factor of two for all the models.
%
In Fig.~\ref{fig:force-field-uncertainty}, 
the BD model is not considered as the predicted positron kinetic energies in this model are too low to be observed at TOA, which is consistent with the definition of the force field in Eq.~(\ref{force-field}).

\begin{figure}[tbp]
    \centering
    \includegraphics[width=12cm]{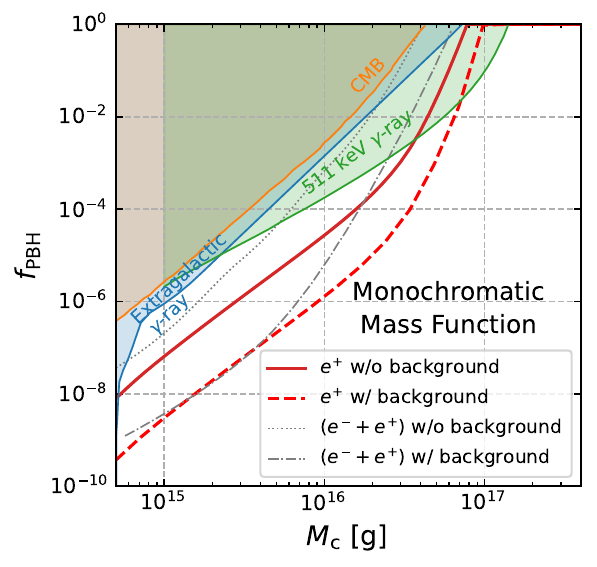}
    \caption{Exclusion lines of PBHs abundance $f_\text{PBH}$. Shadow regions are obtained from different observable, including extragalactic $\gamma$-rays~\cite{Carr:2009jm}, 511 keV $\gamma$-rays data~\cite{Keith:2021guq}, CMB~\cite{Auffinger:2022khh}. 
    The 511 keV constraint is derived from a generalized NFW profile incorporating an additional parameter, specifically an inner slope of $\gamma = 1.6$.
    The dotted and dot-dashed lines are the exclusion lines originally obtained in Ref.~\cite{Boudaud:2018hqb}. The red-solid and red-dashed lines are the results of this work with the employments of model GH and monochromatic mass function, corresponding to the case of including and neglecting CR background, respectively.}
    \label{fig:state_of_art_result}
\end{figure}

In \fig{fig:state_of_art_result}, we compare the constraints from the AMS-02 positron data obtained in the GH model with a selection of constraints from other observables such as that from the extragalactic $\gamma$-ray~\cite{Carr:2009jm}, CMB~\cite{Auffinger:2022khh}, 511\,keV $\gamma$-ray data~\cite{Keith:2021guq}. 
The results in~\cite{Keith:2021guq} considered latitude profile of the 511\,keV emission and assumed a NFW DM density profile with an inner slop $\gamma=1.6$. 
An alternative analysis using longitude emission profile to set constraints on $f_\text{PBH}$ can be found in~\cite{luque2024refininggalacticprimordialblack}.
%

In \fig{fig:state_of_art_result}, we also compare our results with that previously obtained from the Voyager-1 all-electron data in  a diffusive-reacceleration propagation model (model A) in Ref.~\cite{Boudaud:2018hqb}.
The figure shows that the constraints from AMS-02 positron data are competitive with all these previously obtained constraints, and  more stringent than that from the Voyager all-electron data in both the cases with and without including the astrophysical backgrounds.

%

\section{Conclusions}\label{sec:conclusion}
In summary, in this work we have explored the possibility of using the low-energy CR positron data to constrain the abundance of primordial black holes. The advantage of using the CR positron data is based on its secondary origin and thus sensitive to any exotic contributions. We have shown that in some typical diffusive re-acceleration CR propagation models,
the current AMS-02 data can place stringent constraints on $f_{\text{PBH}}$. 
As an example, we have shown that in the {\tt Galprop+Helmod} model for CR propagation in the Galaxy and heliosphere, a conservative upper limit of $f_{\text{PBH}}\lesssim2.15\times 10^{-4}$ at $M_{\mathrm{c}}=2\times 10^{16}$~g can be obtained, which improves the previous constraints from the Voyager-1 data of all-electrons by around an order of magnitude.
Admittedly, the main results of this work is obtained for  diffusive re-acceleration models with  typical $V_a\sim 30~\text{km/s}$ which is supported by a number of analysis~\cite{Trotta:2010mx,Jin:2014ica,Yuan:2018lmc,Boschini:2017fxq,Boschini:2019gow,Boschini:2020jty,DeLaTorreLuque:2021yfq,Luque:2021nxb}, but not yet conclusive.
Compared with other observables such as diffusive $\gamma$-rays,
using the CR flux data to constrain PBH abundance in general suffers from  uncertainties in the propagation models, which are expected to be improved in the future with more precise CR data.
%
For models without re-acceleration, the AMS-02 constraints are not applicable as shown in Fig.~\ref{fig:exclusion_lines_wo_background}, which underscores the necessity of complementary probes such as gamma-ray surveys and synchrotron radiations to constrain PBH scenarios.
\\

\noindent \textbf{Note added:} After submitting the manuscript to arXiv, we were aware of a similar paper \\ arXiv:2403.04988. Their major conclusions are in agreement with this work.

\begin{acknowledgments}
	This work is supported in part by  the NSFC under Grants
	No.~11825506, 
	No.~11821505, and
	No.~12441504.
\end{acknowledgments}

\bibliographystyle{arxivref.bst}
\bibliography{PBH}

\providecommand{\href}[2]{#2}\begingroup\raggedright\begin{thebibliography}{100}

\bibitem{Bauer:2017qwy}
M.~Bauer and T.~Plehn, \href{http://dx.doi.org/10.1007/978-3-030-16234-4}{{\em {Yet Another Introduction to Dark Matter}: {The Particle Physics Approach}}}, vol.~959 of {\em Lecture Notes in Physics}.
\newblock Springer, 2019.
\newblock \href{http://arxiv.org/abs/1705.01987}{{\ttfamily arXiv:1705.01987 [hep-ph]}}.

\bibitem{Abazajian:2012ys}
K.~N. Abazajian {\em et~al.}, ``{Light Sterile Neutrinos: A White Paper},'' \href{http://arxiv.org/abs/1204.5379}{{\ttfamily arXiv:1204.5379 [hep-ph]}}.

\bibitem{Marsh:2015xka}
D.~J.~E. Marsh, ``{Axion Cosmology},'' \href{http://dx.doi.org/10.1016/j.physrep.2016.06.005}{{\em Phys. Rept.} {\bfseries  643} (2016) 1--79}, \href{http://arxiv.org/abs/1510.07633}{{\ttfamily arXiv:1510.07633 [astro-ph.CO]}}.

\bibitem{Hawking:1971ei}
S.~Hawking, ``{Gravitationally collapsed objects of very low mass},'' \href{http://dx.doi.org/10.1093/mnras/152.1.75}{{\em Mon. Not. Roy. Astron. Soc.} {\bfseries  152} (1971) 75}.

\bibitem{Carr:1974nx}
B.~J. Carr and S.~W. Hawking, ``{Black holes in the early Universe},'' \href{http://dx.doi.org/10.1093/mnras/168.2.399}{{\em Mon. Not. Roy. Astron. Soc.} {\bfseries  168} (1974) 399--415}.

\bibitem{Tashiro:2008sf}
H.~Tashiro and N.~Sugiyama, ``{Constraints on Primordial Black Holes by Distortions of Cosmic Microwave Background},'' \href{http://dx.doi.org/10.1103/PhysRevD.78.023004}{{\em Phys. Rev. D} {\bfseries  78} (2008) 023004}, \href{http://arxiv.org/abs/0801.3172}{{\ttfamily arXiv:0801.3172 [astro-ph]}}.

\bibitem{Germani:2018jgr}
C.~Germani and I.~Musco, ``{Abundance of Primordial Black Holes Depends on the Shape of the Inflationary Power Spectrum},'' \href{http://dx.doi.org/10.1103/PhysRevLett.122.141302}{{\em Phys. Rev. Lett.} {\bfseries  122} no.~14, (2019) 141302}, \href{http://arxiv.org/abs/1805.04087}{{\ttfamily arXiv:1805.04087 [astro-ph.CO]}}.

\bibitem{Kannike:2017bxn}
K.~Kannike, L.~Marzola, M.~Raidal, and H.~Veerm\"ae, ``{Single Field Double Inflation and Primordial Black Holes},'' \href{http://dx.doi.org/10.1088/1475-7516/2017/09/020}{{\em JCAP} {\bfseries  09} (2017) 020}, \href{http://arxiv.org/abs/1705.06225}{{\ttfamily arXiv:1705.06225 [astro-ph.CO]}}.

\bibitem{Carr:2017jsz}
B.~Carr, M.~Raidal, T.~Tenkanen, V.~Vaskonen, and H.~Veerm\"ae, ``{Primordial black hole constraints for extended mass functions},'' \href{http://dx.doi.org/10.1103/PhysRevD.96.023514}{{\em Phys. Rev. D} {\bfseries  96} no.~2, (2017) 023514}, \href{http://arxiv.org/abs/1705.05567}{{\ttfamily arXiv:1705.05567 [astro-ph.CO]}}.

\bibitem{Carr:2017edp}
B.~Carr, T.~Tenkanen, and V.~Vaskonen, ``{Primordial black holes from inflaton and spectator field perturbations in a matter-dominated era},'' \href{http://dx.doi.org/10.1103/PhysRevD.96.063507}{{\em Phys. Rev. D} {\bfseries  96} no.~6, (2017) 063507}, \href{http://arxiv.org/abs/1706.03746}{{\ttfamily arXiv:1706.03746 [astro-ph.CO]}}.

\bibitem{MACHO:2000qbb}
{\bfseries  MACHO} Collaboration, C.~Alcock {\em et~al.}, ``{The MACHO project: Microlensing results from 5.7 years of LMC observations},'' \href{http://dx.doi.org/10.1086/309512}{{\em Astrophys. J.} {\bfseries  542} (2000) 281--307}, \href{http://arxiv.org/abs/astro-ph/0001272}{{\ttfamily arXiv:astro-ph/0001272}}.

\bibitem{Wyrzykowski:2010mh}
L.~Wyrzykowski {\em et~al.}, ``{The OGLE View of Microlensing towards the Magellanic Clouds. III. Ruling out sub-solar MACHOs with the OGLE-III LMC data},'' \href{http://dx.doi.org/10.1111/j.1365-2966.2010.18150.x}{{\em Mon. Not. Roy. Astron. Soc.} {\bfseries  413} (2011) 493}, \href{http://arxiv.org/abs/1012.1154}{{\ttfamily arXiv:1012.1154 [astro-ph.GA]}}.

\bibitem{Wyrzykowski:2011tr}
L.~Wyrzykowski {\em et~al.}, ``{The OGLE View of Microlensing towards the Magellanic Clouds. IV. OGLE-III SMC Data and Final Conclusions on MACHOs},'' \href{http://dx.doi.org/10.1111/j.1365-2966.2011.19243.x}{{\em Mon. Not. Roy. Astron. Soc.} {\bfseries  416} (2011) 2949}, \href{http://arxiv.org/abs/1106.2925}{{\ttfamily arXiv:1106.2925 [astro-ph.GA]}}.

\bibitem{Macho:2000nvd}
{\bfseries  Macho} Collaboration, R.~A. Allsman {\em et~al.}, ``{MACHO project limits on black hole dark matter in the 1-30 solar mass range},'' \href{http://dx.doi.org/10.1086/319636}{{\em Astrophys. J. Lett.} {\bfseries  550} (2001) L169}, \href{http://arxiv.org/abs/astro-ph/0011506}{{\ttfamily arXiv:astro-ph/0011506}}.

\bibitem{Griest:2013aaa}
K.~Griest, A.~M. Cieplak, and M.~J. Lehner, ``{Experimental Limits on Primordial Black Hole Dark Matter from the First 2 yr of Kepler Data},'' \href{http://dx.doi.org/10.1088/0004-637X/786/2/158}{{\em Astrophys. J.} {\bfseries  786} no.~2, (2014) 158}, \href{http://arxiv.org/abs/1307.5798}{{\ttfamily arXiv:1307.5798 [astro-ph.CO]}}.

\bibitem{CalchiNovati:2013jpj}
S.~Calchi~Novati, S.~Mirzoyan, P.~Jetzer, and G.~Scarpetta, ``{Microlensing towards the SMC: a new analysis of OGLE and EROS results},'' \href{http://dx.doi.org/10.1093/mnras/stt1402}{{\em Mon. Not. Roy. Astron. Soc.} {\bfseries  435} (2013) 1582}, \href{http://arxiv.org/abs/1308.4281}{{\ttfamily arXiv:1308.4281 [astro-ph.GA]}}.

\bibitem{Griest:2013esa}
K.~Griest, A.~M. Cieplak, and M.~J. Lehner, ``{New Limits on Primordial Black Hole Dark Matter from an Analysis of Kepler Source Microlensing Data},'' \href{http://dx.doi.org/10.1103/PhysRevLett.111.181302}{{\em Phys. Rev. Lett.} {\bfseries  111} no.~18, (2013) 181302}.

\bibitem{Koushiappas:2017chw}
S.~M. Koushiappas and A.~Loeb, ``{Dynamics of Dwarf Galaxies Disfavor Stellar-Mass Black Holes as Dark Matter},'' \href{http://dx.doi.org/10.1103/PhysRevLett.119.041102}{{\em Phys. Rev. Lett.} {\bfseries  119} no.~4, (2017) 041102}, \href{http://arxiv.org/abs/1704.01668}{{\ttfamily arXiv:1704.01668 [astro-ph.GA]}}.

\bibitem{Monroy-Rodriguez:2014ula}
M.~A. Monroy-Rodr\'\i{}guez and C.~Allen, ``{The end of the MACHO era- revisited: new limits on MACHO masses from halo wide binaries},'' \href{http://dx.doi.org/10.1088/0004-637X/790/2/159}{{\em Astrophys. J.} {\bfseries  790} no.~2, (2014) 159}, \href{http://arxiv.org/abs/1406.5169}{{\ttfamily arXiv:1406.5169 [astro-ph.GA]}}.

\bibitem{Carr:2019bel}
B.~Carr, ``{Primordial black holes as dark matter and generators of cosmic structure},'' \href{http://dx.doi.org/10.1007/978-3-030-31593-1_4}{{\em Astrophys. Space Sci. Proc.} {\bfseries  56} (2019) 29--39}, \href{http://arxiv.org/abs/1901.07803}{{\ttfamily arXiv:1901.07803 [astro-ph.CO]}}.

\bibitem{MUSE:2020qbo}
{\bfseries  MUSE} Collaboration, S.~L. Zoutendijk {\em et~al.}, ``{The MUSE-Faint survey: I. Spectroscopic evidence for a star cluster in Eridanus 2 and constraints on MACHOs as a constituent of dark matter},'' \href{http://dx.doi.org/10.1051/0004-6361/201936155}{{\em Astron. Astrophys.} {\bfseries  635} (2020) A107}, \href{http://arxiv.org/abs/2001.08790}{{\ttfamily arXiv:2001.08790 [astro-ph.GA]}}.

\bibitem{Carr:2020gox}
B.~Carr, K.~Kohri, Y.~Sendouda, and J.~Yokoyama, ``{Constraints on primordial black holes},'' \href{http://dx.doi.org/10.1088/1361-6633/ac1e31}{{\em Rept. Prog. Phys.} {\bfseries  84} no.~11, (2021) 116902}, \href{http://arxiv.org/abs/2002.12778}{{\ttfamily arXiv:2002.12778 [astro-ph.CO]}}.

\bibitem{Carr:2020xqk}
B.~Carr and F.~Kuhnel, ``{Primordial Black Holes as Dark Matter: Recent Developments},'' \href{http://dx.doi.org/10.1146/annurev-nucl-050520-125911}{{\em Ann. Rev. Nucl. Part. Sci.} {\bfseries  70} (2020) 355--394}, \href{http://arxiv.org/abs/2006.02838}{{\ttfamily arXiv:2006.02838 [astro-ph.CO]}}.

\bibitem{Hawking:1974rv}
S.~W. Hawking, ``{Black hole explosions},'' \href{http://dx.doi.org/10.1038/248030a0}{{\em Nature} {\bfseries  248} (1974) 30--31}.

\bibitem{Baldes:2020nuv}
I.~Baldes, Q.~Decant, D.~C. Hooper, and L.~Lopez-Honorez, ``{Non-Cold Dark Matter from Primordial Black Hole Evaporation},'' \href{http://dx.doi.org/10.1088/1475-7516/2020/08/045}{{\em JCAP} {\bfseries  08} (2020) 045}, \href{http://arxiv.org/abs/2004.14773}{{\ttfamily arXiv:2004.14773 [astro-ph.CO]}}.

\bibitem{Cheek:2021odj}
A.~Cheek, L.~Heurtier, Y.~F. Perez-Gonzalez, and J.~Turner, ``{Primordial black hole evaporation and dark matter production. I. Solely Hawking radiation},'' \href{http://dx.doi.org/10.1103/PhysRevD.105.015022}{{\em Phys. Rev. D} {\bfseries  105} no.~1, (2022) 015022}, \href{http://arxiv.org/abs/2107.00013}{{\ttfamily arXiv:2107.00013 [hep-ph]}}.

\bibitem{MacGibbon:1991tj}
J.~H. MacGibbon, ``{Quark and gluon jet emission from primordial black holes. 2. The Lifetime emission},'' \href{http://dx.doi.org/10.1103/PhysRevD.44.376}{{\em Phys. Rev. D} {\bfseries  44} (1991) 376--392}.

\bibitem{Carr:2009jm}
B.~J. Carr, K.~Kohri, Y.~Sendouda, and J.~Yokoyama, ``{New cosmological constraints on primordial black holes},'' \href{http://dx.doi.org/10.1103/PhysRevD.81.104019}{{\em Phys. Rev. D} {\bfseries  81} (2010) 104019}, \href{http://arxiv.org/abs/0912.5297}{{\ttfamily arXiv:0912.5297 [astro-ph.CO]}}.

\bibitem{Carr:2016hva}
B.~J. Carr, K.~Kohri, Y.~Sendouda, and J.~Yokoyama, ``{Constraints on primordial black holes from the Galactic gamma-ray background},'' \href{http://dx.doi.org/10.1103/PhysRevD.94.044029}{{\em Phys. Rev. D} {\bfseries  94} no.~4, (2016) 044029}, \href{http://arxiv.org/abs/1604.05349}{{\ttfamily arXiv:1604.05349 [astro-ph.CO]}}.

\bibitem{Arbey:2019vqx}
A.~Arbey, J.~Auffinger, and J.~Silk, ``{Constraining primordial black hole masses with the isotropic gamma ray background},'' \href{http://dx.doi.org/10.1103/PhysRevD.101.023010}{{\em Phys. Rev. D} {\bfseries  101} no.~2, (2020) 023010}, \href{http://arxiv.org/abs/1906.04750}{{\ttfamily arXiv:1906.04750 [astro-ph.CO]}}.

\bibitem{Laha:2020ivk}
R.~Laha, J.~B. Mu\~noz, and T.~R. Slatyer, ``{INTEGRAL constraints on primordial black holes and particle dark matter},'' \href{http://dx.doi.org/10.1103/PhysRevD.101.123514}{{\em Phys. Rev. D} {\bfseries  101} no.~12, (2020) 123514}, \href{http://arxiv.org/abs/2004.00627}{{\ttfamily arXiv:2004.00627 [astro-ph.CO]}}.

\bibitem{Auffinger:2022khh}
J.~Auffinger, ``{Primordial black hole constraints with Hawking radiation\textemdash{}A review},'' \href{http://dx.doi.org/10.1016/j.ppnp.2023.104040}{{\em Prog. Part. Nucl. Phys.} {\bfseries  131} (2023) 104040}, \href{http://arxiv.org/abs/2206.02672}{{\ttfamily arXiv:2206.02672 [astro-ph.CO]}}.

\bibitem{Dasgupta:2019cae}
B.~Dasgupta, R.~Laha, and A.~Ray, ``{Neutrino and positron constraints on spinning primordial black hole dark matter},'' \href{http://dx.doi.org/10.1103/PhysRevLett.125.101101}{{\em Phys. Rev. Lett.} {\bfseries  125} no.~10, (2020) 101101}, \href{http://arxiv.org/abs/1912.01014}{{\ttfamily arXiv:1912.01014 [hep-ph]}}.

\bibitem{Wang:2020uvi}
S.~Wang, D.-M. Xia, X.~Zhang, S.~Zhou, and Z.~Chang, ``{Constraining primordial black holes as dark matter at JUNO},'' \href{http://dx.doi.org/10.1103/PhysRevD.103.043010}{{\em Phys. Rev. D} {\bfseries  103} no.~4, (2021) 043010}, \href{http://arxiv.org/abs/2010.16053}{{\ttfamily arXiv:2010.16053 [hep-ph]}}.

\bibitem{Bernal:2022swt}
N.~Bernal, V.~Mu\~noz Albornoz, S.~Palomares-Ruiz, and P.~Villanueva-Domingo, ``{Current and future neutrino limits on the abundance of primordial black holes},'' \href{http://dx.doi.org/10.1088/1475-7516/2022/10/068}{{\em JCAP} {\bfseries  10} (2022) 068}, \href{http://arxiv.org/abs/2203.14979}{{\ttfamily arXiv:2203.14979 [hep-ph]}}.

\bibitem{Liu:2023cqs}
Q.~Liu and K.~C.~Y. Ng, ``{The Sensitivity Floor for Primordial Black Holes with Neutrino Searches},'' \href{http://arxiv.org/abs/2312.06108}{{\ttfamily arXiv:2312.06108 [hep-ph]}}.

\bibitem{Boudaud:2018hqb}
M.~Boudaud and M.~Cirelli, ``{Voyager 1 $e^\pm$ Further Constrain Primordial Black Holes as Dark Matter},'' \href{http://dx.doi.org/10.1103/PhysRevLett.122.041104}{{\em Phys. Rev. Lett.} {\bfseries  122} no.~4, (2019) 041104}, \href{http://arxiv.org/abs/1807.03075}{{\ttfamily arXiv:1807.03075 [astro-ph.HE]}}.

\bibitem{Laha:2019ssq}
R.~Laha, ``{Primordial Black Holes as a Dark Matter Candidate Are Severely Constrained by the Galactic Center 511 keV $\gamma$ -Ray Line},'' \href{http://dx.doi.org/10.1103/PhysRevLett.123.251101}{{\em Phys. Rev. Lett.} {\bfseries  123} no.~25, (2019) 251101}, \href{http://arxiv.org/abs/1906.09994}{{\ttfamily arXiv:1906.09994 [astro-ph.HE]}}.

\bibitem{Keith:2021guq}
C.~Keith and D.~Hooper, ``{511~keV excess and primordial black holes},'' \href{http://dx.doi.org/10.1103/PhysRevD.104.063033}{{\em Phys. Rev. D} {\bfseries  104} no.~6, (2021) 063033}, \href{http://arxiv.org/abs/2103.08611}{{\ttfamily arXiv:2103.08611 [astro-ph.CO]}}.

\bibitem{luque2024refininggalacticprimordialblack}
P.~D. la~Torre~Luque, J.~Koechler, and S.~Balaji, ``Refining galactic primordial black hole evaporation constraints,'' 2024.
\newblock \url{https://arxiv.org/abs/2406.11949}.

\bibitem{Mittal:2021egv}
S.~Mittal, A.~Ray, G.~Kulkarni, and B.~Dasgupta, ``{Constraining primordial black holes as dark matter using the global 21-cm signal with X-ray heating and excess radio background},'' \href{http://dx.doi.org/10.1088/1475-7516/2022/03/030}{{\em JCAP} {\bfseries  03} (2022) 030}, \href{http://arxiv.org/abs/2107.02190}{{\ttfamily arXiv:2107.02190 [astro-ph.CO]}}.

\bibitem{Maki:1995pa}
K.~Maki, T.~Mitsui, and S.~Orito, ``{Local flux of low-energy anti-protons from evaporating primordial black holes},'' \href{http://dx.doi.org/10.1103/PhysRevLett.76.3474}{{\em Phys. Rev. Lett.} {\bfseries  76} (1996) 3474--3477}, \href{http://arxiv.org/abs/astro-ph/9601025}{{\ttfamily arXiv:astro-ph/9601025}}.

\bibitem{Barrau:2001ev}
A.~Barrau, G.~Boudoul, F.~Donato, D.~Maurin, P.~Salati, and R.~Taillet, ``{Anti-protons from primordial black holes},'' \href{http://dx.doi.org/10.1051/0004-6361:20020313}{{\em Astron. Astrophys.} {\bfseries  388} (2002) 676}, \href{http://arxiv.org/abs/astro-ph/0112486}{{\ttfamily arXiv:astro-ph/0112486}}.

\bibitem{Jin:2017iwg}
H.-B. Jin, Y.-L. Wu, and Y.-F. Zhou, ``{Astrophysical background and dark matter implication based on latest AMS-02 data},'' \href{http://dx.doi.org/10.3847/1538-4357/abb01a}{{\em Astrophys. J.} {\bfseries  901} no.~1, (2020) 80}, \href{http://arxiv.org/abs/1701.02213}{{\ttfamily arXiv:1701.02213 [hep-ph]}}.

\bibitem{Berezinsky:1990qxi}
V.~S. Berezinsky, S.~V. Bulanov, V.~A. Dogiel, and V.~S. Ptuskin, {\em {Astrophysics of cosmic rays}}.
\newblock Springer Berlin, Heidelberg, 1990.

\bibitem{Trotta:2010mx}
R.~Trotta, G.~J\'ohannesson, I.~V. Moskalenko, T.~A. Porter, R.~R.~d. Austri, and A.~W. Strong, ``{Constraints on cosmic-ray propagation models from a global Bayesian analysis},'' \href{http://dx.doi.org/10.1088/0004-637X/729/2/106}{{\em Astrophys. J.} {\bfseries  729} (2011) 106}, \href{http://arxiv.org/abs/1011.0037}{{\ttfamily arXiv:1011.0037 [astro-ph.HE]}}.

\bibitem{Jin:2014ica}
H.-B. Jin, Y.-L. Wu, and Y.-F. Zhou, ``{Cosmic ray propagation and dark matter in light of the latest AMS-02 data},'' \href{http://dx.doi.org/10.1088/1475-7516/2015/09/049}{{\em JCAP} {\bfseries  09} (2015) 049}, \href{http://arxiv.org/abs/1410.0171}{{\ttfamily arXiv:1410.0171 [hep-ph]}}.

\bibitem{Yuan:2018lmc}
Q.~Yuan, C.-R. Zhu, X.-J. Bi, and D.-M. Wei, ``{Secondary cosmic-ray nucleus spectra disfavor particle transport in the Galaxy without reacceleration},'' \href{http://dx.doi.org/10.1088/1475-7516/2020/11/027}{{\em JCAP} {\bfseries  11} (2020GV) 027}, \href{http://arxiv.org/abs/1810.03141}{{\ttfamily arXiv:1810.03141 [astro-ph.HE]}}.

\bibitem{Boschini:2017fxq}
M.~J. Boschini {\em et~al.}, ``{Solution of heliospheric propagation: unveiling the local interstellar spectra of cosmic ray species},'' \href{http://dx.doi.org/10.3847/1538-4357/aa6e4f}{{\em Astrophys. J.} {\bfseries  840} no.~2, (2017) 115}, \href{http://arxiv.org/abs/1704.06337}{{\ttfamily arXiv:1704.06337 [astro-ph.HE]}}.

\bibitem{Boschini:2019gow}
M.~J. Boschini {\em et~al.}, ``{Deciphering the local Interstellar spectra of secondary nuclei with GALPROP/HelMod framework and a hint for primary lithium in cosmic rays},'' \href{http://dx.doi.org/10.3847/1538-4357/ab64f1}{{\em Astrophys. J.} {\bfseries  889} (2020) 167}, \href{http://arxiv.org/abs/1911.03108}{{\ttfamily arXiv:1911.03108 [astro-ph.HE]}}.

\bibitem{Boschini:2020jty}
M.~J. Boschini {\em et~al.}, ``{Inference of the Local Interstellar Spectra of Cosmic-Ray Nuclei Z \ensuremath{\leq} 28 with the GalProp\textendash{}HelMod Framework},'' \href{http://dx.doi.org/10.3847/1538-4365/aba901}{{\em Astrophys. J. Suppl.} {\bfseries  250} no.~2, (2020) 27}, \href{http://arxiv.org/abs/2006.01337}{{\ttfamily arXiv:2006.01337 [astro-ph.HE]}}.

\bibitem{DeLaTorreLuque:2021yfq}
P.~De~La Torre~Luque, M.~N. Mazziotta, F.~Loparco, F.~Gargano, and D.~Serini, ``{Implications of current nuclear cross sections on secondary cosmic rays with the upcoming DRAGON2 code},'' \href{http://dx.doi.org/10.1088/1475-7516/2021/03/099}{{\em JCAP} {\bfseries  03} (2021) 099}, \href{http://arxiv.org/abs/2101.01547}{{\ttfamily arXiv:2101.01547 [astro-ph.HE]}}.

\bibitem{Luque:2021nxb}
P.~D. L.~T. Luque, M.~N. Mazziotta, F.~Loparco, F.~Gargano, and D.~Serini, ``{Markov chain Monte Carlo analyses of the flux ratios of B, Be and Li with the DRAGON2 code},'' \href{http://dx.doi.org/10.1088/1475-7516/2021/07/010}{{\em JCAP} {\bfseries  07} (2021) 010}, \href{http://arxiv.org/abs/2102.13238}{{\ttfamily arXiv:2102.13238 [astro-ph.HE]}}.

\bibitem{DelaTorreLuque:2023huu}
P.~De~la Torre~Luque, S.~Balaji, and P.~Carenza, ``{Multimessenger search for electrophilic feebly interacting particles from supernovae},'' \href{http://dx.doi.org/10.1103/PhysRevD.109.103028}{{\em Phys. Rev. D} {\bfseries  109} no.~10, (2024) 103028}, \href{http://arxiv.org/abs/2307.13731}{{\ttfamily arXiv:2307.13731 [hep-ph]}}.

\bibitem{DelaTorreLuque:2023nhh}
P.~De~la Torre~Luque, S.~Balaji, and P.~Carenza, ``{Robust constraints on feebly interacting particles using XMM-Newton},'' \href{http://dx.doi.org/10.1103/PhysRevD.109.L101305}{{\em Phys. Rev. D} {\bfseries  109} no.~10, (2024) L101305}, \href{http://arxiv.org/abs/2307.13728}{{\ttfamily arXiv:2307.13728 [hep-ph]}}.

\bibitem{DelaTorreLuque:2023olp}
P.~De~la Torre~Luque, S.~Balaji, and J.~Koechler, ``{Importance of Cosmic-Ray Propagation on Sub-GeV Dark Matter Constraints},'' \href{http://dx.doi.org/10.3847/1538-4357/ad41e0}{{\em Astrophys. J.} {\bfseries  968} no.~1, (2024) 46}, \href{http://arxiv.org/abs/2311.04979}{{\ttfamily arXiv:2311.04979 [hep-ph]}}.

\bibitem{DelaTorreLuque:2023cef}
P.~De~la Torre~Luque, S.~Balaji, and J.~Silk, ``{New 511 keV line data provides strongest sub-GeV dark matter constraints},'' \href{http://arxiv.org/abs/2312.04907}{{\ttfamily arXiv:2312.04907 [hep-ph]}}.

\bibitem{Strong:1998pw}
A.~W. Strong and I.~V. Moskalenko, ``{Propagation of cosmic-ray nucleons in the galaxy},'' \href{http://dx.doi.org/10.1086/306470}{{\em Astrophys. J.} {\bfseries  509} (1998) 212--228}, \href{http://arxiv.org/abs/astro-ph/9807150}{{\ttfamily arXiv:astro-ph/9807150}}.

\bibitem{Moskalenko:2001ya}
I.~V. Moskalenko, A.~W. Strong, J.~F. Ormes, and M.~S. Potgieter, ``{Secondary anti-protons and propagation of cosmic rays in the galaxy and heliosphere},'' \href{http://dx.doi.org/10.1086/324402}{{\em Astrophys. J.} {\bfseries  565} (2002) 280--296}, \href{http://arxiv.org/abs/astro-ph/0106567}{{\ttfamily arXiv:astro-ph/0106567}}.

\bibitem{Strong:2001fu}
A.~W. Strong and I.~V. Moskalenko, ``{Models for galactic cosmic ray propagation},'' \href{http://dx.doi.org/10.1016/S0273-1177(01)00112-0}{{\em Adv. Space Res.} {\bfseries  27} (2001) 717--726}, \href{http://arxiv.org/abs/astro-ph/0101068}{{\ttfamily arXiv:astro-ph/0101068}}.

\bibitem{Moskalenko:2002yx}
I.~V. Moskalenko, A.~W. Strong, S.~G. Mashnik, and J.~F. Ormes, ``{Challenging cosmic ray propagation with antiprotons. Evidence for a fresh nuclei component?},'' \href{http://dx.doi.org/10.1086/367697}{{\em Astrophys. J.} {\bfseries  586} (2003) 1050--1066}, \href{http://arxiv.org/abs/astro-ph/0210480}{{\ttfamily arXiv:astro-ph/0210480}}.

\bibitem{Ptuskin:2005ax}
V.~S. Ptuskin, I.~V. Moskalenko, F.~C. Jones, A.~W. Strong, and V.~N. Zirakashvili, ``{Dissipation of magnetohydrodynamic waves on energetic particles: impact on interstellar turbulence and cosmic ray transport},'' \href{http://dx.doi.org/10.1086/501117}{{\em Astrophys. J.} {\bfseries  642} (2006) 902--916}, \href{http://arxiv.org/abs/astro-ph/0510335}{{\ttfamily arXiv:astro-ph/0510335}}.

\bibitem{Bobik:2011ig}
P.~Bobik {\em et~al.}, ``{Systematic Investigation of Solar Modulation of Galactic Protons for Solar Cycle 23 using a Monte Carlo Approach with Particle Drift Effects and Latitudinal Dependence},'' \href{http://dx.doi.org/10.1088/0004-637X/745/2/132}{{\em Astrophys. J.} {\bfseries  745} (2012) 132}, \href{http://arxiv.org/abs/1110.4315}{{\ttfamily arXiv:1110.4315 [astro-ph.SR]}}.

\bibitem{Bobik:2016}
P.~Bobik, M.~J. Boschini, S.~Della~Torre, M.~Gervasi, D.~Grandi, G.~La~Vacca, S.~Pensotti, M.~Putis, P.~G. Rancoita, D.~Rozza, M.~Tacconi, and M.~Zannoni, ``On the forward-backward-in-time approach for monte carlo solution of parker's transport equation: One-dimensional case,'' \href{http://dx.doi.org/https://doi.org/10.1002/2015JA022237}{{\em Journal of Geophysical Research: Space Physics} {\bfseries  121} no.~5, (2016) 3920--3930}.

\bibitem{Boschini:2017gic}
M.~J. Boschini, S.~Della~Torre, M.~Gervasi, G.~La~Vacca, and P.~G. Rancoita, ``{Propagation of cosmic rays in heliosphere: The HELMOD model},'' \href{http://dx.doi.org/10.1016/j.asr.2017.04.017}{{\em Adv. Space Res.} {\bfseries  62} (2018) 2859--2879}, \href{http://arxiv.org/abs/1704.03733}{{\ttfamily arXiv:1704.03733 [astro-ph.SR]}}.

\bibitem{Boschini:2019ubh}
M.~J. Boschini, S.~Della~Torre, M.~Gervasi, G.~La~Vacca, and P.~G. Rancoita, ``{The HelMod Model in the Works for Inner and Outer Heliosphere: from AMS to Voyager Probes Observations},'' \href{http://dx.doi.org/10.1016/j.asr.2019.04.007}{{\em Adv. Space Res.} {\bfseries  64} no.~12, (2019) 2459--2476}, \href{http://arxiv.org/abs/1903.07501}{{\ttfamily arXiv:1903.07501 [physics.space-ph]}}.

\bibitem{Boschini:2022jwz}
M.~J. Boschini, S.~Della~Torre, M.~Gervasi, G.~La~Vacca, and P.~G. Rancoita, ``{Forecasting of cosmic rays intensities with HelMod Model},'' \href{http://dx.doi.org/10.1016/j.asr.2022.01.031}{{\em Adv. Space Res.} {\bfseries  70} no.~9, (2022) 2649--2657}.

\bibitem{Auffinger:2022sqj}
J.~Auffinger and A.~Arbey, ``{Beyond the Standard Model with BlackHawk v2.0},'' \href{http://dx.doi.org/10.22323/1.409.0017}{{\em PoS} {\bfseries  CompTools2021} (2022) 017}, \href{http://arxiv.org/abs/2207.03266}{{\ttfamily arXiv:2207.03266 [gr-qc]}}.

\bibitem{DeLuca:2019buf}
V.~De~Luca, V.~Desjacques, G.~Franciolini, A.~Malhotra, and A.~Riotto, ``{The initial spin probability distribution of primordial black holes},'' \href{http://dx.doi.org/10.1088/1475-7516/2019/05/018}{{\em JCAP} {\bfseries  05} (2019) 018}, \href{http://arxiv.org/abs/1903.01179}{{\ttfamily arXiv:1903.01179 [astro-ph.CO]}}.

\bibitem{1704.06573}
T.~Chiba and S.~Yokoyama, ``{Spin Distribution of Primordial Black Holes},'' \href{http://dx.doi.org/10.1093/ptep/ptx087}{{\em PTEP} {\bfseries  2017} no.~8, (2017) 083E01}, \href{http://arxiv.org/abs/1704.06573}{{\ttfamily arXiv:1704.06573 [gr-qc]}}.

\bibitem{1901.05963}
M.~Mirbabayi, A.~Gruzinov, and J.~Nore\~na, ``{Spin of Primordial Black Holes},'' \href{http://dx.doi.org/10.1088/1475-7516/2020/03/017}{{\em JCAP} {\bfseries  03} (2020) 017}, \href{http://arxiv.org/abs/1901.05963}{{\ttfamily arXiv:1901.05963 [astro-ph.CO]}}.

\bibitem{Harada:2016mhb}
T.~Harada, C.-M. Yoo, K.~Kohri, K.-i. Nakao, and S.~Jhingan, ``{Primordial black hole formation in the matter-dominated phase of the Universe},'' \href{http://dx.doi.org/10.3847/1538-4357/833/1/61}{{\em Astrophys. J.} {\bfseries  833} no.~1, (2016) 61}, \href{http://arxiv.org/abs/1609.01588}{{\ttfamily arXiv:1609.01588 [astro-ph.CO]}}.

\bibitem{Harada:2017fjm}
T.~Harada, C.-M. Yoo, K.~Kohri, and K.-I. Nakao, ``{Spins of primordial black holes formed in the matter-dominated phase of the Universe},'' \href{http://dx.doi.org/10.1103/PhysRevD.96.083517}{{\em Phys. Rev. D} {\bfseries  96} no.~8, (2017) 083517}, \href{http://arxiv.org/abs/1707.03595}{{\ttfamily arXiv:1707.03595 [gr-qc]}}. [Erratum: Phys.Rev.D 99, 069904 (2019)].

\bibitem{Cotner:2017tir}
E.~Cotner and A.~Kusenko, ``{Primordial black holes from scalar field evolution in the early universe},'' \href{http://dx.doi.org/10.1103/PhysRevD.96.103002}{{\em Phys. Rev. D} {\bfseries  96} no.~10, (2017) 103002}, \href{http://arxiv.org/abs/1706.09003}{{\ttfamily arXiv:1706.09003 [astro-ph.CO]}}.

\bibitem{Hawking:1975iha}
S.~W. Hawking, ``{Particle Creation by Black Holes},'' in {\em {1st Oxford Conference on Quantum Gravity}}.
\newblock 8, 1975.

\bibitem{Page:1976df}
D.~N. Page, ``{Particle Emission Rates from a Black Hole: Massless Particles from an Uncharged, Nonrotating Hole},'' \href{http://dx.doi.org/10.1103/PhysRevD.13.198}{{\em Phys. Rev. D} {\bfseries  13} (1976) 198--206}.

\bibitem{Arbey:2019mbc}
A.~Arbey and J.~Auffinger, ``{BlackHawk: A public code for calculating the Hawking evaporation spectra of any black hole distribution},'' \href{http://dx.doi.org/10.1140/epjc/s10052-019-7161-1}{{\em Eur. Phys. J. C} {\bfseries  79} no.~8, (2019) 693}, \href{http://arxiv.org/abs/1905.04268}{{\ttfamily arXiv:1905.04268 [gr-qc]}}.

\bibitem{1907.11846}
A.~Coogan, L.~Morrison, and S.~Profumo, ``{Hazma: A Python Toolkit for Studying Indirect Detection of Sub-GeV Dark Matter},'' \href{http://dx.doi.org/10.1088/1475-7516/2020/01/056}{{\em JCAP} {\bfseries  01} (2020) 056}, \href{http://arxiv.org/abs/1907.11846}{{\ttfamily arXiv:1907.11846 [hep-ph]}}.

\bibitem{Dolgov:1992pu}
A.~Dolgov and J.~Silk, ``{Baryon isocurvature fluctuations at small scales and baryonic dark matter},'' \href{http://dx.doi.org/10.1103/PhysRevD.47.4244}{{\em Phys. Rev. D} {\bfseries  47} (1993) 4244--4255}.

\bibitem{Clesse:2015wea}
S.~Clesse and J.~Garc\'\i{}a-Bellido, ``{Massive Primordial Black Holes from Hybrid Inflation as Dark Matter and the seeds of Galaxies},'' \href{http://dx.doi.org/10.1103/PhysRevD.92.023524}{{\em Phys. Rev. D} {\bfseries  92} no.~2, (2015) 023524}, \href{http://arxiv.org/abs/1501.07565}{{\ttfamily arXiv:1501.07565 [astro-ph.CO]}}.

\bibitem{MacGibbon:1990zk}
J.~H. MacGibbon and B.~R. Webber, ``{Quark and gluon jet emission from primordial black holes: The instantaneous spectra},'' \href{http://dx.doi.org/10.1103/PhysRevD.41.3052}{{\em Phys. Rev. D} {\bfseries  41} (1990) 3052--3079}.

\bibitem{Chao:2021orr}
W.~Chao, T.~Li, and J.~Liao, ``{Connecting Primordial Black Hole to boosted sub-GeV Dark Matter through neutrino},'' \href{http://arxiv.org/abs/2108.05608}{{\ttfamily arXiv:2108.05608 [hep-ph]}}.

\bibitem{Webber:1992dks}
W.~R. Webber, M.~A. Lee, and M.~Gupta, ``{Propagation of cosmic-ray nuclei in a diffusing galaxy with convective halo and thin matter disk},'' \href{http://dx.doi.org/10.1086/171262}{{\em Astrophys. J.} {\bfseries  390} (1992) 96}.

\bibitem{Strong:2007nh}
A.~W. Strong, I.~V. Moskalenko, and V.~S. Ptuskin, ``{Cosmic-ray propagation and interactions in the Galaxy},'' \href{http://dx.doi.org/10.1146/annurev.nucl.57.090506.123011}{{\em Ann. Rev. Nucl. Part. Sci.} {\bfseries  57} (2007) 285--327}, \href{http://arxiv.org/abs/astro-ph/0701517}{{\ttfamily arXiv:astro-ph/0701517}}.

\bibitem{1941DoSSR..30..301K}
A.~{Kolmogorov}, ``{The Local Structure of Turbulence in Incompressible Viscous Fluid for Very Large Reynolds' Numbers},'' {\em Akademiia Nauk SSSR Doklady} {\bfseries  30} (Jan., 1941) 301--305.

\bibitem{1964SvA.....7..566I}
P.~S. {Iroshnikov}, ``{Turbulence of a Conducting Fluid in a Strong Magnetic Field},'' {\em Soviet Astronomy} {\bfseries  7} (Feb., 1964) 566.

\bibitem{10.1063/1.1761412}
R.~H. Kraichnan, ``{Inertial‐Range Spectrum of Hydromagnetic Turbulence},'' \href{http://dx.doi.org/10.1063/1.1761412}{{\em The Physics of Fluids} {\bfseries  8} no.~7, (07, 1965) 1385--1387}, \href{http://arxiv.org/abs/https://pubs.aip.org/aip/pfl/article-pdf/8/7/1385/12617184/1385\_1\_online.pdf}{{\ttfamily https://pubs.aip.org/aip/pfl/article-pdf/8/7/1385/12617184/1385\_1\_online.pdf}}. \url{https://doi.org/10.1063/1.1761412}.

\bibitem{1994ApJ...431..705S}
E.~S. {Seo} and V.~S. {Ptuskin}, ``{Stochastic Reacceleration of Cosmic Rays in the Interstellar Medium},'' \href{http://dx.doi.org/10.1086/174520}{{\em The Astrophysical Journal} {\bfseries  431} (Aug., 1994) 705}.

\bibitem{Maurin:2002ua}
D.~Maurin, R.~Taillet, F.~Donato, P.~Salati, A.~Barrau, and G.~Boudoul, ``{Galactic cosmic ray nuclei as a tool for astroparticle physics},'' \href{http://arxiv.org/abs/astro-ph/0212111}{{\ttfamily arXiv:astro-ph/0212111}}.

\bibitem{Moskalenko:1997gh}
I.~V. Moskalenko and A.~W. Strong, ``{Production and propagation of cosmic ray positrons and electrons},'' \href{http://dx.doi.org/10.1086/305152}{{\em Astrophys. J.} {\bfseries  493} (1998) 694--707}, \href{http://arxiv.org/abs/astro-ph/9710124}{{\ttfamily arXiv:astro-ph/9710124}}.

\bibitem{1976ApJ...208..346G}
M.~A. {Gordon} and W.~B. {Burton}, ``{Carbon monoxide in the Galaxy. I. The radial distribution of CO, H$_{2}$, and nucleons.},'' \href{http://dx.doi.org/10.1086/154613}{{\em \apj} {\bfseries  208} (Sept., 1976) 346--353}.

\bibitem{1986A&A...155..380C}
P.~{Cox}, E.~{Kruegel}, and P.~G. {Mezger}, ``{Principal heating sources of dust in the galactic disk.},'' {\em Astronomy and Astrophysics} {\bfseries  155} (Feb., 1986) 380--396.

\bibitem{1988ApJ...324..248B}
L.~{Bronfman}, R.~S. {Cohen}, H.~{Alvarez}, J.~{May}, and P.~{Thaddeus}, ``{A CO Survey of the Southern Milky Way: The Mean Radial Distribution of Molecular Clouds within the Solar Circle},'' \href{http://dx.doi.org/10.1086/165892}{{\em \apj} {\bfseries  324} (Jan., 1988) 248}.

\bibitem{1991Natur.354..121C}
J.~M. {Cordes}, J.~M. {Weisberg}, D.~A. {Frail}, S.~R. {Spangler}, and M.~{Ryan}, ``{The galactic distribution of free electrons},'' \href{http://dx.doi.org/10.1038/354121a0}{{\em \nat} {\bfseries  354} no.~6349, (Nov., 1991) 121--124}.

\bibitem{1997ApJ...480..173S}
T.~J. {Sodroski}, N.~{Odegard}, R.~G. {Arendt}, E.~{Dwek}, J.~L. {Weiland}, M.~G. {Hauser}, and T.~{Kelsall}, ``{A Three-dimensional Decomposition of the Infrared Emission from Dust in the Milky Way},'' \href{http://dx.doi.org/10.1086/303961}{{\em Astrophsical Journal} {\bfseries  480} no.~1, (May, 1997) 173--187}.

\bibitem{Kappl:2015hxv}
R.~Kappl, ``{SOLARPROP: Charge-sign Dependent Solar Modulation for Everyone},'' \href{http://dx.doi.org/10.1016/j.cpc.2016.05.025}{{\em Comput. Phys. Commun.} {\bfseries  207} (2016) 386--399}, \href{http://arxiv.org/abs/1511.07875}{{\ttfamily arXiv:1511.07875 [astro-ph.SR]}}.

\bibitem{Vittino:2017fuh}
A.~Vittino, C.~Evoli, and D.~Gaggero, ``{Cosmic-ray transport in the heliosphere with HelioProp},'' \href{http://dx.doi.org/10.22323/1.301.0024}{{\em PoS} {\bfseries  ICRC2017} (2018) 024}, \href{http://arxiv.org/abs/1707.09003}{{\ttfamily arXiv:1707.09003 [astro-ph.HE]}}.

\bibitem{Gleeson:1968zza}
L.~J. Gleeson and W.~I. Axford, ``{Solar Modulation of Galactic Cosmic Rays},'' \href{http://dx.doi.org/10.1086/149822}{{\em Astrophys. J.} {\bfseries  154} (1968) 1011}.

\bibitem{Simon:1996dk}
M.~Simon and U.~Heinbach, ``{Production of anti-protons in interstellar space by propagating cosmic rays under conditions of diffusive reacceleration},'' \href{http://dx.doi.org/10.1086/176676}{{\em Astrophys. J.} {\bfseries  456} (1996) 519--524}.

\bibitem{Johannesson:2016rlh}
G.~J\'ohannesson {\em et~al.}, ``{Bayesian analysis of cosmic-ray propagation: evidence against homogeneous diffusion},'' \href{http://dx.doi.org/10.3847/0004-637X/824/1/16}{{\em Astrophys. J.} {\bfseries  824} no.~1, (2016) 16}, \href{http://arxiv.org/abs/1602.02243}{{\ttfamily arXiv:1602.02243 [astro-ph.HE]}}.

\bibitem{Korsmeier:2021brc}
M.~Korsmeier and A.~Cuoco, ``{Implications of Lithium to Oxygen AMS-02 spectra on our understanding of cosmic-ray diffusion},'' \href{http://dx.doi.org/10.1103/PhysRevD.103.103016}{{\em Phys. Rev. D} {\bfseries  103} no.~10, (2021) 103016}, \href{http://arxiv.org/abs/2103.09824}{{\ttfamily arXiv:2103.09824 [astro-ph.HE]}}.

\bibitem{Genolini:2019ewc}
Y.~G\'enolini {\em et~al.}, ``{Cosmic-ray transport from AMS-02 boron to carbon ratio data: Benchmark models and interpretation},'' \href{http://dx.doi.org/10.1103/PhysRevD.99.123028}{{\em Phys. Rev. D} {\bfseries  99} no.~12, (2019) 123028}, \href{http://arxiv.org/abs/1904.08917}{{\ttfamily arXiv:1904.08917 [astro-ph.HE]}}.

\bibitem{Weinrich:2020cmw}
N.~Weinrich, Y.~G\'enolini, M.~Boudaud, L.~Derome, and D.~Maurin, ``{Combined analysis of AMS-02 (Li,Be,B)/C, N/O, 3He, and 4He data},'' \href{http://dx.doi.org/10.1051/0004-6361/202037875}{{\em Astron. Astrophys.} {\bfseries  639} (2020) A131}, \href{http://arxiv.org/abs/2002.11406}{{\ttfamily arXiv:2002.11406 [astro-ph.HE]}}.

\bibitem{Orlando:2013ysa}
E.~Orlando and A.~Strong, ``{Galactic synchrotron emission with cosmic ray propagation models},'' \href{http://dx.doi.org/10.1093/mnras/stt1718}{{\em Mon. Not. Roy. Astron. Soc.} {\bfseries  436} (2013) 2127}, \href{http://arxiv.org/abs/1309.2947}{{\ttfamily arXiv:1309.2947 [astro-ph.GA]}}.

\bibitem{Neronov:2011wi}
A.~Neronov, D.~V. Semikoz, and A.~M. Taylor, ``{Low-energy break in the spectrum of Galactic cosmic rays},'' \href{http://dx.doi.org/10.1103/PhysRevLett.108.051105}{{\em Phys. Rev. Lett.} {\bfseries  108} (2012) 051105}, \href{http://arxiv.org/abs/1112.5541}{{\ttfamily arXiv:1112.5541 [astro-ph.HE]}}.

\bibitem{Boschini:2018zdv}
M.~J. Boschini {\em et~al.}, ``{HelMod in the works: from direct observations to the local interstellar spectrum of cosmic-ray electrons},'' \href{http://dx.doi.org/10.3847/1538-4357/aaa75e}{{\em Astrophys. J.} {\bfseries  854} no.~2, (2018) 94}, \href{http://arxiv.org/abs/1801.04059}{{\ttfamily arXiv:1801.04059 [astro-ph.HE]}}.

\bibitem{AMS:2023anq}
{\bfseries  AMS} Collaboration, M.~Aguilar {\em et~al.}, ``{Properties of Cosmic-Ray Sulfur and Determination of the Composition of Primary Cosmic-Ray Carbon, Neon, Magnesium, and Sulfur: Ten-Year Results from the Alpha Magnetic Spectrometer},'' \href{http://dx.doi.org/10.1103/PhysRevLett.130.211002}{{\em Phys. Rev. Lett.} {\bfseries  130} no.~21, (2023) 211002}.

\bibitem{Navarro:1996gj}
J.~F. Navarro, C.~S. Frenk, and S.~D.~M. White, ``{A Universal density profile from hierarchical clustering},'' \href{http://dx.doi.org/10.1086/304888}{{\em Astrophys. J.} {\bfseries  490} (1997) 493--508}, \href{http://arxiv.org/abs/astro-ph/9611107}{{\ttfamily arXiv:astro-ph/9611107}}.

\bibitem{Stone:2019}
E.~Stone, A.~Cummings, and B.~Heikkila, ``Cosmic ray measurements from voyager 2 as it crossed into interstellar space,'' \href{http://dx.doi.org/10.1038/s41550-019-0928-3}{{\em Nature Astronomy} {\bfseries  3} (11, 2019) 1013--1018}.

\bibitem{AMS:2021nhj}
{\bfseries  AMS} Collaboration, M.~Aguilar {\em et~al.}, ``{The Alpha Magnetic Spectrometer (AMS) on the international space station: Part II \textemdash{} Results from the first seven years},'' \href{http://dx.doi.org/10.1016/j.physrep.2020.09.003}{{\em Phys. Rept.} {\bfseries  894} (2021) 1--116}.

\bibitem{Orusa:2022pvp}
L.~Orusa, M.~Di~Mauro, F.~Donato, and M.~Korsmeier, ``{New determination of the production cross section for secondary positrons and electrons in the Galaxy},'' \href{http://dx.doi.org/10.1103/PhysRevD.105.123021}{{\em Phys. Rev. D} {\bfseries  105} no.~12, (2022) 123021}, \href{http://arxiv.org/abs/2203.13143}{{\ttfamily arXiv:2203.13143 [astro-ph.HE]}}.

\bibitem{Kafexhiu:2014cua}
E.~Kafexhiu, F.~Aharonian, A.~M. Taylor, and G.~S. Vila, ``{Parametrization of gamma-ray production cross-sections for pp interactions in a broad proton energy range from the kinematic threshold to PeV energies},'' \href{http://dx.doi.org/10.1103/PhysRevD.90.123014}{{\em Phys. Rev. D} {\bfseries  90} no.~12, (2014) 123014}, \href{http://arxiv.org/abs/1406.7369}{{\ttfamily arXiv:1406.7369 [astro-ph.HE]}}.

\bibitem{Kamae:2006bf}
T.~Kamae, N.~Karlsson, T.~Mizuno, T.~Abe, and T.~Koi, ``{Parameterization of Gamma, e+/- and Neutrino Spectra Produced by p-p Interaction in Astronomical Environment},'' \href{http://dx.doi.org/10.1086/513602}{{\em Astrophys. J.} {\bfseries  647} (2006) 692--708}, \href{http://arxiv.org/abs/astro-ph/0605581}{{\ttfamily arXiv:astro-ph/0605581}}. [Erratum: Astrophys.J. 662, 779 (2007)].

\bibitem{DelaTorreLuque:2023zyd}
P.~De~la Torre~Luque, F.~Loparco, and M.~N. Mazziotta, ``{The FLUKA cross sections for cosmic-ray leptons and uncertainties on current positron predictions},'' \href{http://dx.doi.org/10.1088/1475-7516/2023/10/011}{{\em JCAP} {\bfseries  10} (2023) 011}, \href{http://arxiv.org/abs/2305.02958}{{\ttfamily arXiv:2305.02958 [astro-ph.HE]}}.

\bibitem{Koldobskiy:2021nld}
S.~Koldobskiy, M.~Kachelrie\ss{}, A.~Lskavyan, A.~Neronov, S.~Ostapchenko, and D.~V. Semikoz, ``{Energy spectra of secondaries in proton-proton interactions},'' \href{http://dx.doi.org/10.1103/PhysRevD.104.123027}{{\em Phys. Rev. D} {\bfseries  104} no.~12, (2021) 123027}, \href{http://arxiv.org/abs/2110.00496}{{\ttfamily arXiv:2110.00496 [astro-ph.HE]}}.

\bibitem{Bierlich:2022pfr}
C.~Bierlich {\em et~al.}, ``{A comprehensive guide to the physics and usage of PYTHIA 8.3},'' \href{http://dx.doi.org/10.21468/SciPostPhysCodeb.8}{{\em SciPost Phys. Codeb.} {\bfseries  2022} (2022) 8}, \href{http://arxiv.org/abs/2203.11601}{{\ttfamily arXiv:2203.11601 [hep-ph]}}.

\bibitem{DiMauro:2023oqx}
M.~Di~Mauro, F.~Donato, M.~Korsmeier, S.~Manconi, and L.~Orusa, ``{Novel prediction for secondary positrons and electrons in the Galaxy},'' \href{http://dx.doi.org/10.1103/PhysRevD.108.063024}{{\em Phys. Rev. D} {\bfseries  108} no.~6, (2023) 063024}, \href{http://arxiv.org/abs/2304.01261}{{\ttfamily arXiv:2304.01261 [astro-ph.HE]}}.

\end{thebibliography}\endgroup
\newpage
\appendix
\end{document}